\documentclass[preprint,12pt,authoryear]{elsarticle}
\usepackage{amssymb}
\usepackage{amsthm}
\usepackage{textcomp}
\usepackage{subcaption}
\usepackage{amsmath}
\usepackage{breakurl} 
\usepackage[hyphens]{url} 
\usepackage[hidelinks]{hyperref} 

\journal{Int. J. Heat Fluid Flow}

\begin{document}

\begin{frontmatter}



\title{Deep-reinforcement-learning-based separation control in a two-dimensional airfoil}

\author{Xavier Garcia$^a$}
\affiliation{FLOW, Engineering Mechanics, KTH Royal Institute of Technology, Stockholm, Sweden}
\author{Arnau Miró$^{c,b}$}
\affiliation{Barcelona Supercomputing Center, Barcelona, Spain}
\affiliation{Universitat Politecnica de Catalunya, Barcelona, Spain}
\author{Pol Suárez$^a$}

\author{Francisco Alcántara-Ávila$^a$}

\author{Jean Rabault$^d$}
\affiliation{Independent Researcher, Oslo, Norway}
\author{Bernat Font$^e$}
\affiliation{Faculty of Mechanical Engineering, Technische Universiteit Delft, The Netherlands}

\author{Oriol Lehmkuhl$^b$}

\author{Ricardo Vinuesa$^a$}

\date{\today}



\begin{abstract}
 The aim of this study is to discover new active-flow-control (AFC) techniques for separation mitigation in a two-dimensional NACA 0012 airfoil at a Reynolds number of 3000. To find these AFC strategies, a framework consisting of a deep-reinforcement-learning (DRL) agent has been used to determine the action strategies to apply to the flow. The actions involve blowing and suction through jets at the airfoil surface. The flow is simulated with the numerical code Alya, which is a low-dissipation finite-element code, on a high-performance computing system.
Various control strategies obtained through DRL led to 43.9\% drag reduction, while others yielded an increase in aerodynamic efficiency of 58.6\%. In comparison, periodic-control strategies demonstrated lower energy efficiency while failing to achieve the same level of aerodynamic improvements as the DRL-based approach. These gains have been attained through the implementation of a dynamic, closed-loop, time-dependent, active control mechanism over the airfoil.
\end{abstract}







\end{frontmatter}


\section{\label{sec:introduction}Introduction}

Active flow control (AFC) research is crucial not only for the development of aerodynamics and aircraft design but also for all transport-related systems. This is primarily due to the critical relationship between drag reduction and overall system efficiency. Flow control strategies, whether passive (implemented through geometric modifications or additions) or active (utilizing devices that dynamically alter the flow over time), enable the manipulation of flow behavior to enhance aerodynamic performance. 

Active flow control facilitates key aerodynamic benefits such as drag reduction, fuel consumption minimization, and noise reduction. Additionally, AFC can enhance lift, improve maneuverability—particularly in low-speed vehicles—and even contribute to radar signature reduction by strategically disturbing the flow.

Over the past decades, various passive flow control techniques have been developed. The earliest advancements in this field, including those during World War II, are reviewed by \citet{mclellan1988history}. A notable example is the surface-roughness technique studied by \citet{beratlis_separation_2017}, where different surface textures alter the boundary layer, delaying flow separation and thereby reducing drag. Another significant contribution is the work of \citet{bechert_viscous_1989}, who investigated the aerodynamic effects of fine longitudinal ribs in viscous flows. 

Following the exploration of passive methods, the field moved toward active techniques. The first significant approach involved periodic excitations, as demonstrated by \citet{Amitay1998}. Later, \citet{greenblatt_control_2000} extended these efforts to prevent flow separation using similar excitation strategies. A major breakthrough was the implementation of closed-loop controllers, as demonstrated by \citet{muddada_active_2010}, who used simple actuators to reduce drag in a low Reynolds number (\(Re\)) cylinder flow, achieving a $53\%$ drag reduction.

More relevant to the present study, blowing and suction techniques have been widely explored for boundary layer control. \citet{kametani_direct_2011} investigated their effects on flow stability, while \citet{voevodin_improvement_2019} and \citet{yousefi_three-dimensional_2015} studied suction and ejection mechanisms for drag reduction and flow control optimization.

In recent years, modern AFC strategies have increasingly incorporated deep reinforcement learning (DRL), utilizing artificial neural networks (ANNs) for autonomous control strategy discovery. DRL has been recognized for its ability to optimize flow control problems in a computationally efficient manner. The first demonstration of DRL in AFC was provided by \citet{rabault_artificial_2019}, where a DRL agent controlled the 2D flow around a cylinder at \(Re = 100\), achieving an 8\% drag reduction using a proximal-policy-optimization (PPO) algorithm. Later, \citet{rabault_accelerating_2019} enhanced the framework through parallelization, significantly accelerating training efficiency. \citet{tang_robust_2020} extended DRL-based AFC to higher Reynolds numbers using a more complex setup involving four synthetic jets.

Building on these early advancements, an increasing amount of research has confirmed the effectiveness of DRL for AFC applications. \citet{ren_applying_2021} explored weakly turbulent 2D cylinder flows, while \citet{act11120359} expanded these findings to higher Reynolds number cylinders. \citet{cavallazzi_deep_2024} applied DRL to Couette flow, showcasing its versatility across different flow types. \citet{font_deep_2025} further demonstrated DRL’s adaptability in controlling turbulent separation bubbles. Additionally, DRL applications have been extended to Rayleigh-Bénard convection cases, as examined by \citet{vignon_effective_2023} and \citet{vasanth_multi-agent_2024}.

Given the rapid progress in this field, interested readers may refer to broader review articles such as \citet{vignon_recent_2023} for an extensive overview of DRL-based AFC.

Among the most relevant prior studies, \citet{suarez_flow_20243} explored the first DRL-controlled three-dimensional (3D) cylinder flow, addressing transitional turbulent regimes in cylinder wakes by incorporating multiple control jets in the spanwise direction and managing complex flow structures. Building on this, \citet{suarez_active_2024} extended the study to higher Reynolds numbers, where turbulence becomes even more intricate, further increasing the complexity of the active flow control (AFC) problem.

For AFC applications on airfoils, \citet{zong_experimental_2024} applied Q-learning to regulate plasma synthetic jets at an airfoil's trailing edge, aiming to suppress flow separation. Additionally, \citet{wang_deep_2022} explored DRL-based synthetic jet control on a NACA 0012 airfoil at \(Re_D = 3000\), employing three jets on the upper surface. Their results demonstrated a $27\%$ drag reduction and a $27.7\%$ lift enhancement. However, their strategy utilized a static jet intensity, lacking real-time adaptability as they use an open-loop structure.

This study aims to control the recirculation bubble that forms at the trailing edge of a two-dimensional NACA 0012 airfoil using a DRL-based computational framework. The methodology builds upon the approaches of \citet{rabault_artificial_2019}, \citet{suarez_flow_20243}, and \citet{suarez_active_2024}, integrating a proximal-policy-optimization (PPO) algorithm (\citet{schulman_proximal_2017}) for active flow control. The computational experiments are conducted at \(Re_D = 3000\) using three synthetic jets placed on the upper airfoil surface. These jets are dynamically controlled by a DRL agent to optimize aerodynamic performance by reducing drag and increasing aerodynamic efficiency.

\begin{raggedright}
The simulations are executed on the Nord 3 Supercomputer at the Barcelona Supercomputing Center (BSC-CNS) using Alya, a high-performance computational fluid dynamics (CFD) solver optimized for large-scale simulations \citep{vazquez_alya_2014}.
\end{raggedright}

The remainder of this paper is organized as follows. Section~\ref{sec:Methodology} presents the methodology, detailing the deep reinforcement learning (DRL) framework, the numerical setup, and the validation of the Alya solver. Section~\ref{sec:results} discusses the DRL training results for active flow control (AFC) on a two-dimensional airfoil. Finally, Section~\ref{sec:conclusions} summarizes the key findings and provides concluding remarks on the study’s contributions and potential future directions.

\section{\label{sec:Methodology}Methodology}

\subsection{Domain, Mesh, and Numerical Method}
This study investigates DRL-based active flow control in a NACA 0012 airfoil immersed in a two-dimensional channel flow. The selected configuration facilitates both validation and direct comparison with the findings of \citet{wang_deep_2022}. All geometric parameters are normalized with respect to the airfoil chord length ($D$)\footnote{To maintain consistency with \citet{wang_deep_2022}, we use \(D\) to represent the airfoil chord length. This differs from the common convention where \(c\) is used to represent the chord.}, which serves as the reference scale throughout the study. 

\begin{figure}[tbp!]
    \centering
    \includegraphics[trim={23cm 2cm 4cm 2cm},clip,width=0.78\textwidth]{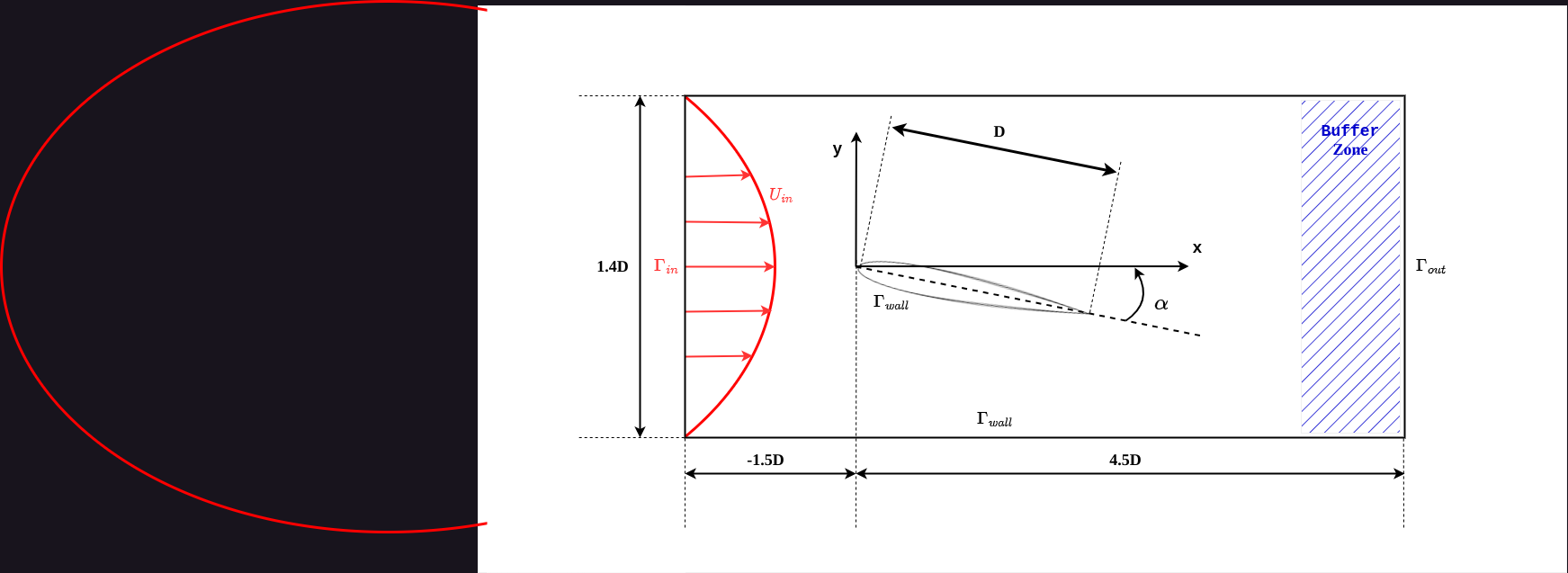}
    \caption{Computational domain dimensions, normalized by the airfoil chord length $D$. The angle of attack $\alpha$ is illustrated, with boundary conditions represented by $\Gamma$. The buffer zone is highlighted in dark blue.}
    \label{fig:domain}
\end{figure}

The computational domain, depicted in Figure~\ref{fig:domain}, extends $6D$ in the streamwise direction and $1.4D$ in the perpendicular direction ($y$). The coordinate system's origin is located at the leading edge of the airfoil, positioned at $(x = 1.5D, y = 0)$. The airfoil is oriented at an angle of attack of $10^\circ$, achieved through the appropriate rotational transformation.

The inlet boundary condition $\Gamma_{\rm{in}}$ is defined by a parabolic velocity profile, given by

\begin{equation}\label{eq:1}
    U_{\rm{in}}(y) =  \frac{4 U_m (0.7D - y)(0.7D + y)}{1.4D^2},
\end{equation}

where $U_m = 0.45$ represents the maximum velocity at the profile's center. The Reynolds number is defined as

\begin{equation}\label{eq:3}
    Re_D = \frac{\bar{U} D}{\nu},
\end{equation}

where $\nu$ is the kinematic viscosity, and $\bar{U}$ represents the mean velocity, computed as

\begin{equation}\label{eq:2}
    \bar{U} = \int_{-0.7D}^{+0.7D} U_{\text{in}}(y) \,\mathrm{d}y = \frac{2}{3} U_\mathrm{m}.
\end{equation}

From this, the mean velocity at the inlet is determined to be $\bar{U} = 0.3$.

A no-slip condition is applied at the airfoil surface and channel walls $\Gamma_{\rm{wall}}$. At the outlet boundary $\Gamma_{\rm{out}}$, a zero-gradient condition is enforced on velocity, while maintaining a constant pressure.

To enhance numerical stability, a buffer zone is incorporated near the outlet, wherein the fluid viscosity is artificially increased by an order of magnitude. This technique, adapted from \citet{ferziger_computational_2020}, effectively dissipates vortices, prevents recirculation instabilities, and ensures mass conservation. The absence of this stabilization mechanism could lead to numerical divergence and simulation failure due to unphysical recirculation effects at the outflow boundary.

The mesh used in the simulations is an unstructured triangular mesh with refinement near the airfoil, around the control jets, and in the wake region.

Three active control jets are positioned on the upper surface of the airfoil at $x/D = 0.2$, $0.3$, and $0.4$, respectively. The actuators, labeled as $\mathrm{jet}_1$, $\mathrm{jet}_2$, and $\mathrm{jet}_3$, are configured to ensure that actuation occurs perpendicular to the main flow direction. This setup ensures that any observed improvements in aerodynamic performance are due to active control rather than direct streamwise momentum injection. The jets produce a parabolic velocity profile, which ensures zero velocity at both edges.

The mass flow rate profile of each jet is defined by the following equation:

\begin{equation}\label{eq:4}
    Q_i(x) = Q_{a_i} \sin \left( \frac{\pi (x - x_{2_i})}{x_{1_i} - x_{2_i}} \right),
\end{equation}

where $i = 1, 2, 3$ corresponds to the three jets. Here, $Q_i$ represents the mass flow rate profile of each jet, $Q_{a_i}$ is the value selected by the agent for each jet, and $x_{1_i}$ and $x_{2_i}$ denote the initial and final streamwise coordinates of each jet, respectively.

To ensure that blowing or suction always occurs perpendicular to the local airfoil surface, each jet is oriented normally to the surface by applying a rotation angle $\theta$, given by:

\begin{equation}\label{eq:5}
    \theta = 2 \tan^{-1}\left( \frac{x_1 - x_2}{(y_2 - y_1) + \sqrt{(y_2 - y_1)^2 + (x_1 - x_2)^2}} \right).
\end{equation}

Thus, the velocity components in the streamwise and perpendicular directions are expressed as:

\begin{align} \label{eq:66}
    Q_x &= Q  \cos(\theta), \\
    Q_y &= Q  \sin(\theta).
\end{align}

\begin{figure}[htbp!]
    \centering
    \includegraphics[trim={1cm 1cm 0cm 0cm},clip,width=0.6\linewidth]{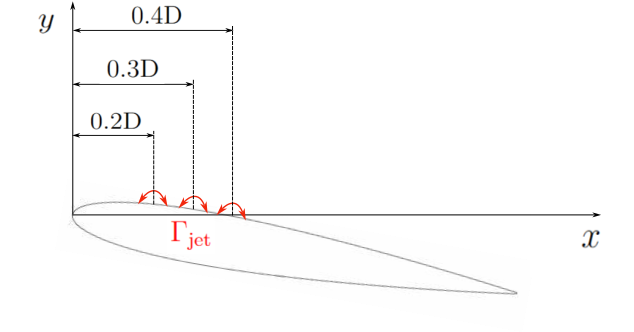}
    \caption{Jets location with respect to the leading edge of the airfoil.}
    \label{fig:airfoil_jets}
\end{figure}

The numerical method used by Alya to compute the forces on the airfoil surface solves the incompressible viscous Navier-Stokes equations. These equations are formulated for the entire domain $\left(\Omega\right)$ as follows:

\begin{align}\label{eq:6}
    \partial_t u + (u \cdot \nabla) u - \nabla \cdot (2 \nu \epsilon) + \nabla p = f, \\
    \nabla \cdot u = 0,
\end{align}

where $u$ represents the velocity field, and $\epsilon$ denotes the velocity strain-rate tensor, which is defined as a function of the velocity $\mathbf{u}$ by

\begin{equation}
    \epsilon = \frac{1}{2} (\nabla \mathbf{u} + \nabla^T \mathbf{u}).
\end{equation}

The term $f$ corresponds to the external body force acting on the fluid.

The convective component in the non-linear term, $C_{\textrm{nonc}}(\mathbf{u}) = (\mathbf{u} \cdot \nabla)\mathbf{u}$, is formulated to ensure conservation of energy, momentum, and angular momentum, as detailed in \citet{charnyi_conservation_2017,charnyi_efficient_2019}.

For time discretization, a semi-implicit Runge-Kutta scheme of second order is employed for the convective term, while a Crank-Nicolson scheme is applied for the diffusive term. During time integration, Alya utilizes an eigenvalue-based time-stepping method, as implemented by \citet{trias_self-adaptive_2011}.

At each time step, the numerical solution of these equations is obtained, and the drag force $(F_D)$ and lift force $(F_L)$ are computed by integrating over the entire surface $(S)$ of the airfoil:

\begin{equation}
    \mathbf{F} = \int (\zeta \cdot \mathbf{n})  \mathbf{e}_j \, \mathrm{d}S,
\end{equation}

where $\zeta$ represents the Cauchy stress tensor, $n$ is the unit normal vector to the surface, and $e_j$ is a unit vector aligned with the main flow velocity for drag computation, and perpendicular to it for lift computation. The aerodynamic coefficients are then determined as:

\begin{eqnarray}
    C_d = \frac{2F_d}{\rho \bar{U}^2 D},\\
    C_l = \frac{2F_l}{\rho \bar{U}^2 D}.
\end{eqnarray}
\subsection{DRL characteristics}\label{sec:DRL_characteristics}

\begin{figure*}
    \centering
    \includegraphics[trim={0cm 0cm 0cm 0cm},clip,width=0.8\linewidth]{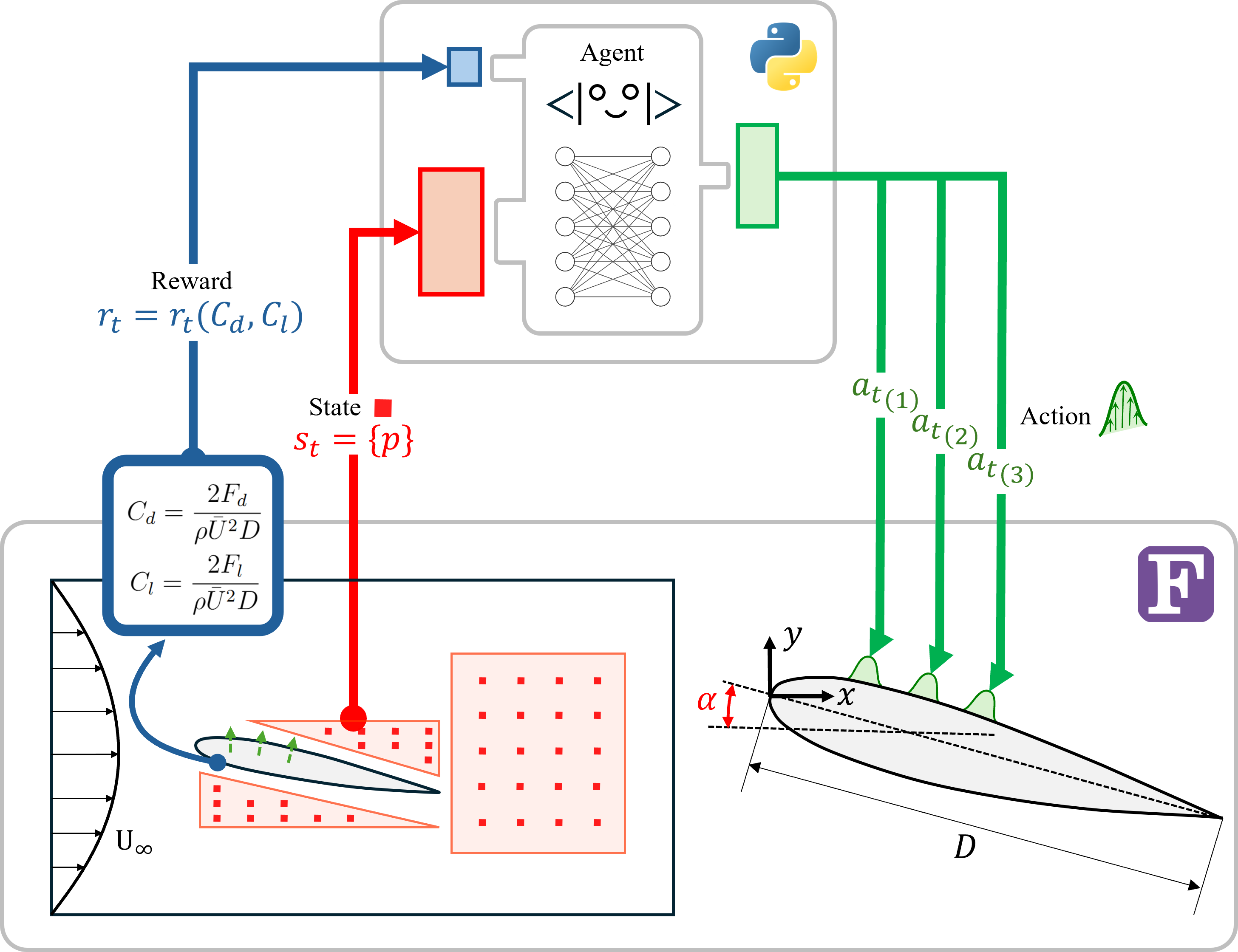}
    \caption{Schematic representation that illustrates the reinforcement-learning framework
 applied to a two-dimensional airfoil, showing communication channels between two main actors.
 In this case, the direction of the information is clockwise. At the top, we show the agent architecture
 featuring a shared neural network. At the bottom, the computational-fluid-dynamics (CFD) environ
ment is depicted, with the airfoil chord $D$ as the reference length}
    \label{fig:drl-framework}
\end{figure*}

The DRL framework is illustrated in Figure \ref{fig:drl-framework}, which depicts each step in the training process. Initially, the solver, Alya, performs a simulation without any jet interaction. This reference simulation, referred to as the baseline, is executed from $0$ to $200$ time units (TU). Once the baseline simulation is completed, the episodes are executed sequentially, with each episode starting from the final step of the baseline.

Each episode consists of a sequence of actions $(a_t)$, corresponding to the actuation of the jets. The selection of these actions is performed by the agent, which makes decisions based on the received state $(s_t)$ and the corresponding reward $(r_t)$.


Once an episode is completed, the next episode begins, utilizing the information from the previous one while also starting from the last time step of the baseline simulation. When all episodes have been simulated, a complete training cycle is concluded.

\begin{figure*}
    \centering
    \includegraphics[trim={0cm 0cm 0cm 0cm},clip,width=0.8\linewidth]{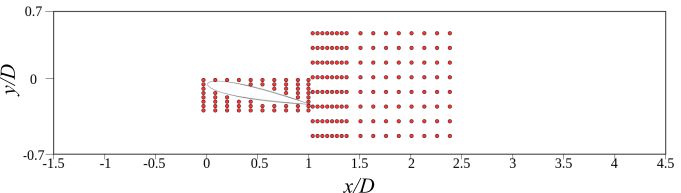}
    \caption{Schematic representation of the computational domain, where red dots indicate the locations of probes used to define the state \( s_t \). The probes are strategically placed around the wing and in the wake region to capture relevant flow information.}
    \label{fig:witnespoints}
\end{figure*}

The DRL environment communicates with the agent through three distinct channels. The first channel is the observation state, $s_t$, which represents the pressure values at various predefined locations in the computational domain, referred to as probes or witness points. These points are strategically placed in critical areas around the airfoil and in the wake region, as illustrated in Figure \ref{fig:witnespoints}. The extracted pressure values are normalized to ensure that the data received by the agent remains within the range $[-1,1]$.

The second communication channel is the action, $a_t$, which corresponds to the value determined by the agent and applied to the control jets for active flow manipulation. Although the values of $a_t$ are initially defined within the range $[-1,1]$ for optimal processing by the DRL algorithm. The intensity of the actuation is further refined using Equations \ref{eq:66}, 7, and \ref{eq:q_de_t}, ensuring a smooth temporal transition. The selected value of \(Q_t\) is determined by the agent based on the policy function $\pi\left(a_t|s_t\right)$.

\begin{equation} \label{eq:q_de_t}
    Q(t) = Q_t + g(t)  (Q_{t+1} - Q_{t}),
\end{equation}

where $t = (t-t_t)/(t_{t+1}-t_t)$, and the function $g(t)$ is defined as

\begin{equation}
    g(t) = \frac{f(t)}{f(t) + f(1-t)},
\end{equation}

with $f(t)$ given by an exponential function that ensures a smooth transition:

\begin{equation}
f(t) = \left\{
\begin{array}{lll}
e^{-\frac{1}{t}}, & \text{if} & t > 0, \\
0, & \text{if} & t \leq 0.
\end{array}
\right.
\end{equation}

Maintaining a smooth transition is crucial to facilitating the learning process for the agent while avoiding abrupt changes that could introduce instability. Previous experiments have demonstrated that using step or linear functions poses challenges for effective learning and computational burden.

The final communication channel between the environment and the agent is the reward, $r_t$, which serves as the primary criterion for the PPO agent to decide future actions. The reward function can be formulated based on different objectives, depending on the specific focus of each training scenario. In this study, three different reward functions are considered. The first function prioritizes drag reduction while maintaining lift stability, as expressed in Equation \ref{eq:reward_1}. The second function maximizes aerodynamic efficiency, as shown in Equation \ref{eq:reward_2}. The third function introduces adjustable weights for lift and drag contributions, providing greater control over the agent's optimization strategy, as represented in Equation \ref{eq:reward_393}.

\begin{equation}\label{eq:reward_1}
    r_1 = \mu  \left( \sigma + ((-C_d + C_{d_0}) - \lambda  (C_l - C_{l_0}) \right),
\end{equation}

\begin{equation}\label{eq:reward_2}
    r_2 = \mu  \left( \frac{C_l}{C_d} + \sigma \right),
\end{equation}

\begin{equation}\label{eq:reward_393}
    r_3 = \mu  \left( w_1  (C_{d_0} - C_d) + w_2  (C_l - C_{l_0}) \right) 
\end{equation}

In these equations, $\mu$ represents the normalization factor, and $\sigma$ is an offset term ensuring that the reward starts at zero at the beginning of each episode. The parameter $\lambda$ serves as a penalization factor for lift variation in the first reward function. The terms $C_d$ and $C_l$ denote the drag and lift coefficients, respectively, averaged up to the current time step. The values $C_{d_0}$ and $C_{l_0}$ correspond to the baseline drag and lift coefficients. In the third reward function, $w_1$ and $w_2$ represent adjustable weight parameters for drag and lift, respectively, allowing for a more tailored optimization approach.

Finally, Table \ref{tab:methodologycharcteristics} provides a summary of all the parameters considered, highlighting the key characteristics of the two main cases examined in this study.

\begin{table}[htbp!]
\centering
\renewcommand{\arraystretch}{1.1} 
\begin{tabular}{lcc}
\hline \hline
\multicolumn{3}{c}{\textbf{Shared Parameters}} \\ \hline
Mesh Cells & \multicolumn{2}{c}{184790} \\ 
Witness Points & \multicolumn{2}{c}{178} \\ 
\(Q_{\rm{norm}}\) & \multicolumn{2}{c}{1.25} \\ 
Time per Action & \multicolumn{2}{c}{0.275} \\ 
Actions per Episode & \multicolumn{2}{c}{300} \\ 
Batch Size & \multicolumn{2}{c}{33} \\ 
CPUs - Environment & \multicolumn{2}{c}{16} \\ 
Environments & \multicolumn{2}{c}{33} \\ 
Total CPUs & \multicolumn{2}{c}{544} \\ 
Network & \multicolumn{2}{c}{\{[Dense, 512], [Dense, 512]\}} \\ 
Update Frequency & \multicolumn{2}{c}{1.0} \\ 
Learning Rate & \multicolumn{2}{c}{0.001} \\ 
Multi-Step & \multicolumn{2}{c}{25} \\ 
Subsampling Fraction & \multicolumn{2}{c}{0.2} \\ 
Likelihood Ratio Clipping & \multicolumn{2}{c}{0.2} \\ 
Discount & \multicolumn{2}{c}{0.99} \\ 
Optimizer Steps & \multicolumn{2}{c}{5} \\ 
Optimizer Type & \multicolumn{2}{c}{Adam} \\ 
Optimizer Learning Rate & \multicolumn{2}{c}{0.001} \\ \hline \hline

\multicolumn{3}{c}{\textbf{Case-Specific Parameters}} \\ \hline

\textbf{Parameter} & \textbf{Case 1} & \textbf{Case 2} \\  
 & \(C_d\) Reduction & \(C_l/C_d\) Enhancement \\ \hline \hline

\(\mu\) & 1.0 & 1.2 \\ 
\(\lambda\) & 0.3 & N/A \\ 
\(\sigma\) & 0.0 & -3.88 \\ 
\(w_1\) & N/A & 0.3 \\ 
\(w_2\) & N/A & 0.7 \\ 
\(Q_{\rm{max}}\) & $\pm$ 0.5 & $\pm$ 1.5 \\ 
Number of Episodes & 957 & 1188 \\ \hline \hline
\end{tabular}
\caption{Comparison of parameters used in the simulations and DRL framework for the two cases: (1) \(C_d\) reduction and (2) \(C_l/C_d\) enhancement. Shared parameters apply to both cases, while case-specific parameters differ between the two configurations.}
\label{tab:methodologycharcteristics}
\end{table}

\subsection{CFD Validation}
The framework has been previously validated for flow control around cylinders in studies such as \citet{act11120359} and \citet{suarez_flow_20243}. To extend this validation to an airfoil configuration, it is necessary to perform a baseline simulation without actuation. This baseline allows for a direct comparison with the results of \citet{wang_deep_2022} in terms of lift and drag coefficients.

\begin{figure}[htbp!]
    \centering
    \begin{subfigure}[b]{0.48\textwidth}
        \centering
        \includegraphics[trim={0cm 0cm 0cm 0cm},clip,width=\linewidth]{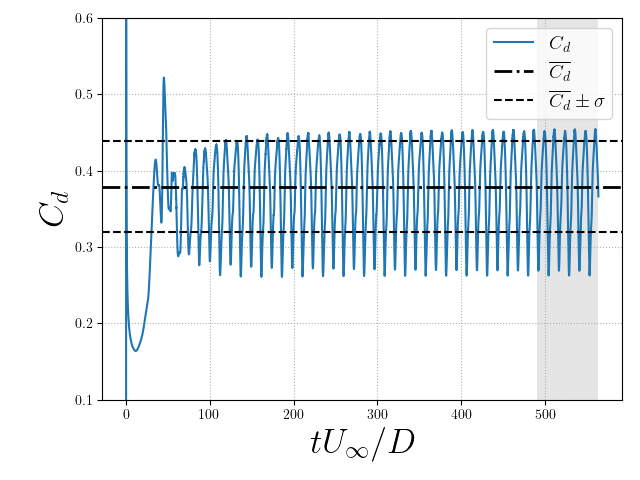}
        \caption{Drag coefficient (\(C_d\)) during the baseline simulation, with its average value (\(\overline{C_d}\)).}
        \label{fig:baselinedrag}
    \end{subfigure}
    \hfill
    \begin{subfigure}[b]{0.48\textwidth}
        \centering
        \includegraphics[trim={0cm 0cm 0cm 0cm},clip,width=\linewidth]{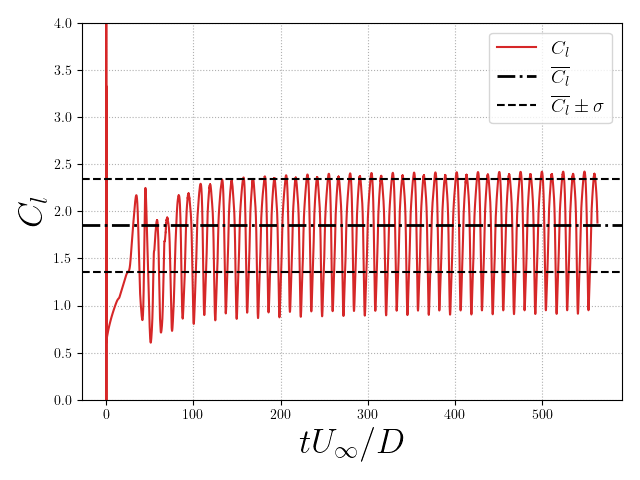}
        \caption{Lift coefficient (\(C_l\)) during the baseline simulation, with its average value (\(\overline{C_l}\)).}
        \label{fig:baselinelift}
    \end{subfigure}
    \caption{Baseline aerodynamic coefficients for drag and lift.}
    \label{fig:baseline_combined}
\end{figure}

The baseline results are presented in Figures \ref{fig:baselinedrag} and \ref{fig:baselinelift}. The computed average drag coefficient is \(C_d = 0.376\), while the average lift coefficient is \(C_l = 1.822\), yielding an aerodynamic efficiency of \(C_l/C_d = 4.846\).

When compared to the results reported by \citet{wang_deep_2022}, the computed drag coefficient exhibits a difference of approximately 15\%. However, a more significant discrepancy is observed in the lift coefficient, where the variation is around 45\%. To ensure the reliability of these results, multiple tests were conducted with different mesh resolutions. Despite these efforts, the extracted results remained consistent, confirming their accuracy. The discrepancy in the aerodynamic drag coefficient can be attributed to differences in mesh resolution and solver characteristics.

\begin{figure}[htbp!]
    \centering
    \begin{subfigure}[b]{0.48\textwidth}
        \centering
        \raisebox{1.5cm}{\includegraphics[trim={2cm 1cm 1cm 2cm},clip,width=\linewidth]{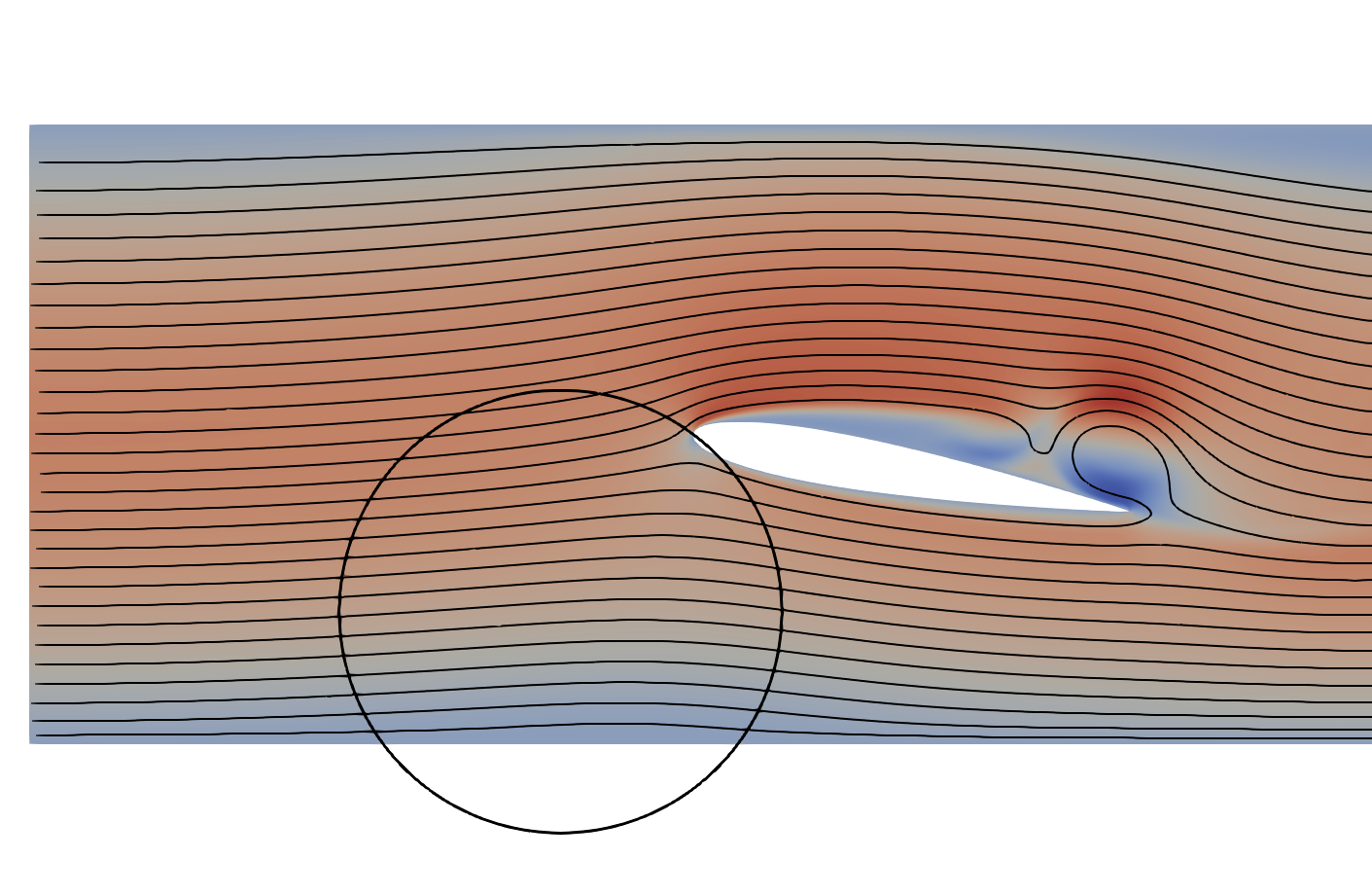}}
        
        \caption{Flow streamlines depicting the ground effect.}
        \label{fig:ge1}
    \end{subfigure}
    \hfill
    \begin{subfigure}[b]{0.48\textwidth}
        \centering
        \raisebox{0cm}{\includegraphics[trim={1cm 0cm 1cm 0.5cm},clip,width=\linewidth]{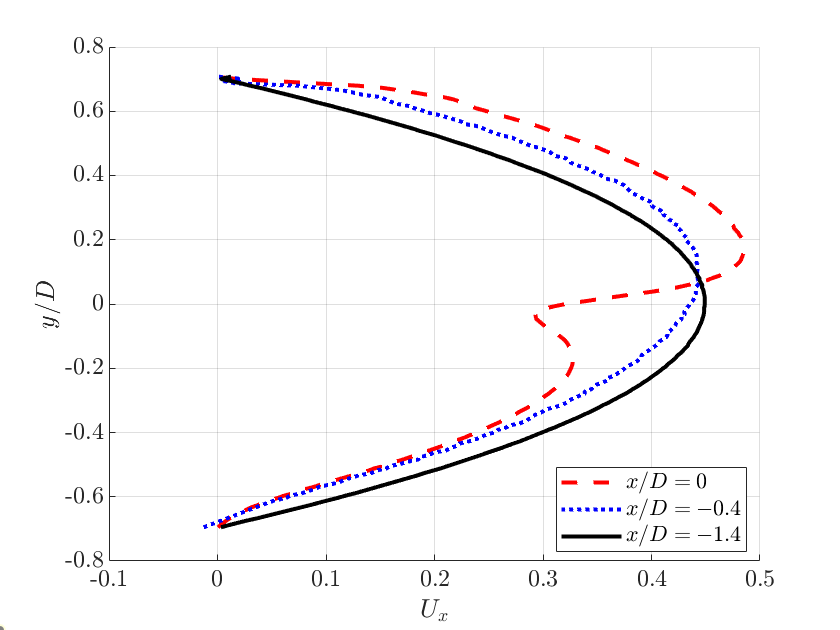}}
        \caption{Velocity distribution in the \(x\)-direction at locations \((x/D = -1.4D, -0.4D, 0D)\), highlighting ground effect asymmetry before interaction with the airfoil.}
        \label{fig:ge2}
    \end{subfigure}
    \caption{Representation of the ground effect: (a) Flow streamlines and (b) velocity profile in the \(x\)-direction at selected locations.}
    \label{fig:ge_combined}
\end{figure}

The influence of ground effect in channel flow configurations has been widely studied in computational fluid dynamics, particularly in the context of aerodynamic performance. The present study aligns with prior research, such as \citet{vinuesa_minimum_2015}, which examines the impact of confined flow environments on aerodynamic forces and vortex dynamics. This effect is evident in Figures \ref{fig:ge1} and \ref{fig:ge2}, where the compression of flow streamlines near the leading edge and beneath the airfoil illustrates the presence of ground effect. Additionally, Figure \ref{fig:ge2} highlights a loss of inflow symmetry before the airfoil interaction, further confirming the influence of confinement on the upstream velocity field.

The displacement of the peak velocity towards a higher region and the modification of the velocity slope within the range \(\left[-0.7, -0.2\right]\) suggest a localized reduction in velocity in the lower region, contributing to an additional lift component that is not directly generated by the airfoil. These observations are consistent with the findings of \citet{vinuesa_minimum_2015}, which emphasize the role of flow confinement in modifying pressure distributions and aerodynamic forces.

Moreover, the present study reinforces the idea that detached vortices in confined environments exhibit increased asymmetry and growth due to altered dissipation properties. The Alya solver, with its enhanced resolution capabilities, captures these flow structures more effectively, resulting in larger and more asymmetric vortices. This is particularly significant when considering the validation process, as running the baseline simulation at \(Re_D = 3000\) in an open-domain configuration (i.e., without channel confinement) produces results that align well with previous studies, including those of \citet{swanson_comparison_2016}. Thus, the findings not only validate the solver’s accuracy but also underline the importance of considering confinement effects when analyzing aerodynamic performance in channel flows.

\section{Results and Discussion}\label{sec:results}

This section is divided into two parts. The first part illustrates the results obtained from intensity-fixed solutions following the same approach as the study conducted by \citet{wang_deep_2022}. The final part discusses the outcomes of applying deep reinforcement learning (DRL) for synthetic jet control in airfoil training.

\subsection{Application of Fixed-Intensity Control in Synthetic Jets}\label{sec:intensityfixed}

This subsection presents the results obtained by applying a fixed-intensity control strategy, following the approach described in \citet{wang_deep_2022}. The primary objective is to validate state-of-the-art techniques and assess variations in intensity magnitude to determine whether different values lead to improved performance.

To establish a direct comparison, it is necessary to translate the mass flow rate values from \citet{wang_deep_2022} into the framework used in this study. Their work employs a dimensionless parameter, \(Q^*\), defined based on a reference mass flow rate, \(Q_{\mathrm{ref}} = 0.3\), which can be expressed in terms of the normal mass flow rate as follows:

\begin{equation}
    Q_{\mathrm{jet}_i} = Q^* \hspace{0.1cm}  Q_{\mathrm{ref}}.
\end{equation}

The numerical framework in this study allows for tuning the values selected by the agent, which must be appropriately translated for use in the computational fluid dynamics (CFD) solver. This transformation is achieved using Equation \ref{eq:translate}:

\begin{equation}\label{eq:translate}
    Q_{\mathrm{jet}_i} = Q_{\mathrm{agent}}  Q(x),
\end{equation}

where \(Q(x)\) is given by:

\begin{equation}\label{eq:integral}
    \resizebox{0.91\textwidth}{!}{$
     Q(x) = \int_{x_1}^{x_2} \sin \left[ \pi  \frac{x-x_2}{x_1 - x_2} \right] 
      \cos \left[  2  \arctan \left( 
     \frac{x_1 - x_2}{(y_2 - y_1) + \sqrt{(y_2 - y_1)^2+(x_1 -x_2)^2}} 
     \right) \right] \, \mathrm{d}x,
    $}
\end{equation}

where \(x_1\) and \(x_2\) correspond to the start and end coordinates of the jet in the \(x\)-direction, while \(y_1\) and \(y_2\) represent the corresponding coordinates in the \(y\)-direction.

The values of \(Q_{\mathrm{agent}}\) corresponding to the same mass flow rate as in \citet{wang_deep_2022} are presented in Table \ref{tab:2}.

\begin{table}[htbp!]
\centering
\renewcommand{\arraystretch}{1.3}
\begin{tabular}{ccc}
\hline \hline
\textbf{\(Q_{\mathrm{jet}_1}\)} & \textbf{\(Q_{\mathrm{jet}_2}\)} & \textbf{\(Q_{\mathrm{jet}_3}\)} \\ \hline
$-1.25$                  & $1.25$                   & $-0.125$                 \\ \hline \hline
\end{tabular}
\caption{Jet intensities used in \citet{wang_deep_2022} translated into this study's framework.}
\label{tab:2}
\end{table}

\begin{figure}[htbp!]
    \centering
    \begin{subfigure}[b]{0.49\linewidth}
        \centering
        \includegraphics[trim={0cm 0.1cm 0cm 0.3cm},clip,width=\linewidth]{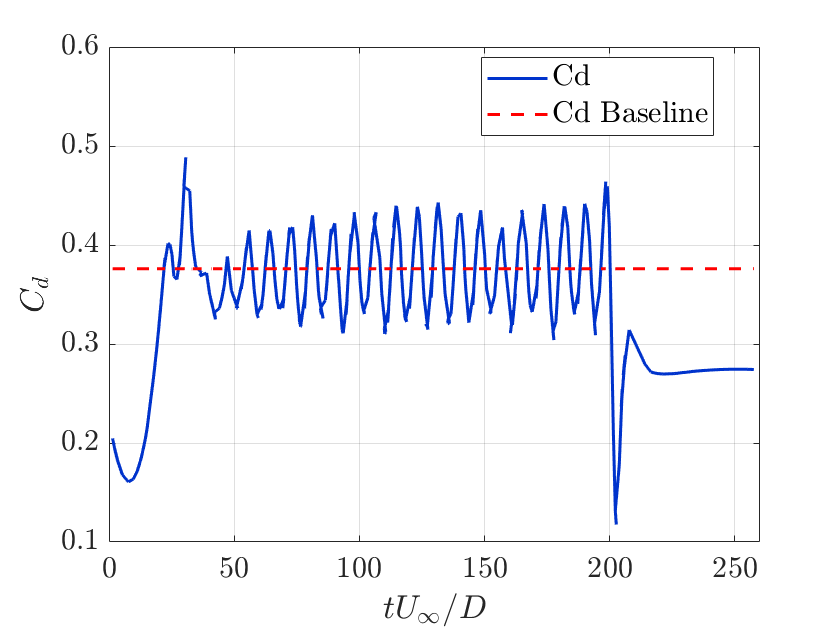}
    \end{subfigure}
    \hfill
    \begin{subfigure}[b]{0.49\linewidth}
        \centering
        \includegraphics[trim={0cm 0.1cm 0cm 0.3cm},clip,width=\linewidth]{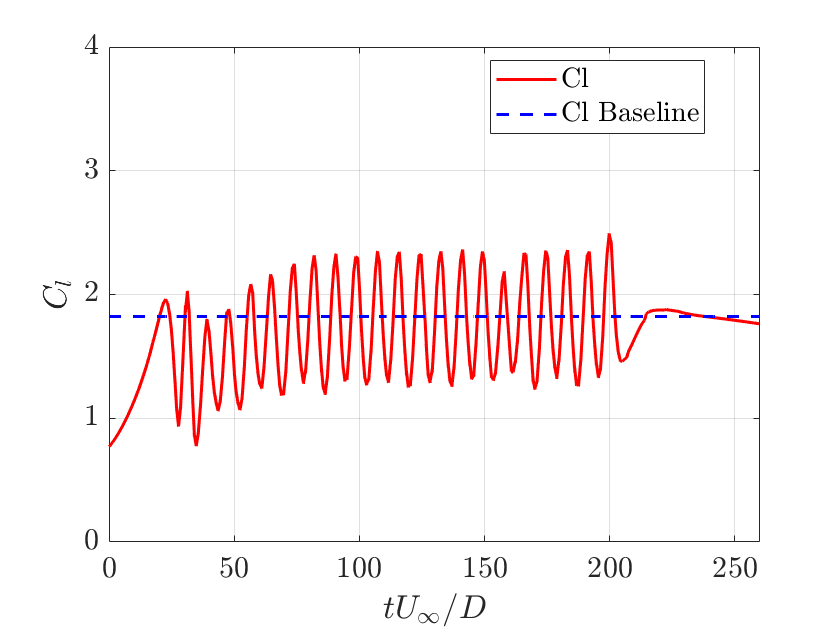}
    \end{subfigure}
    \caption{Evolution of drag and lift coefficients when applying the fixed-intensity strategy from \citet{wang_deep_2022} compared to baseline values.}
    \label{fig:fixed_intensitiy_wang2}
\end{figure}

\begin{figure}[htbp!]
\centering

\begin{subfigure}[b]{1\linewidth}
    \centering
    \includegraphics[trim={0 9cm 0 9cm},clip,width=0.7\linewidth]{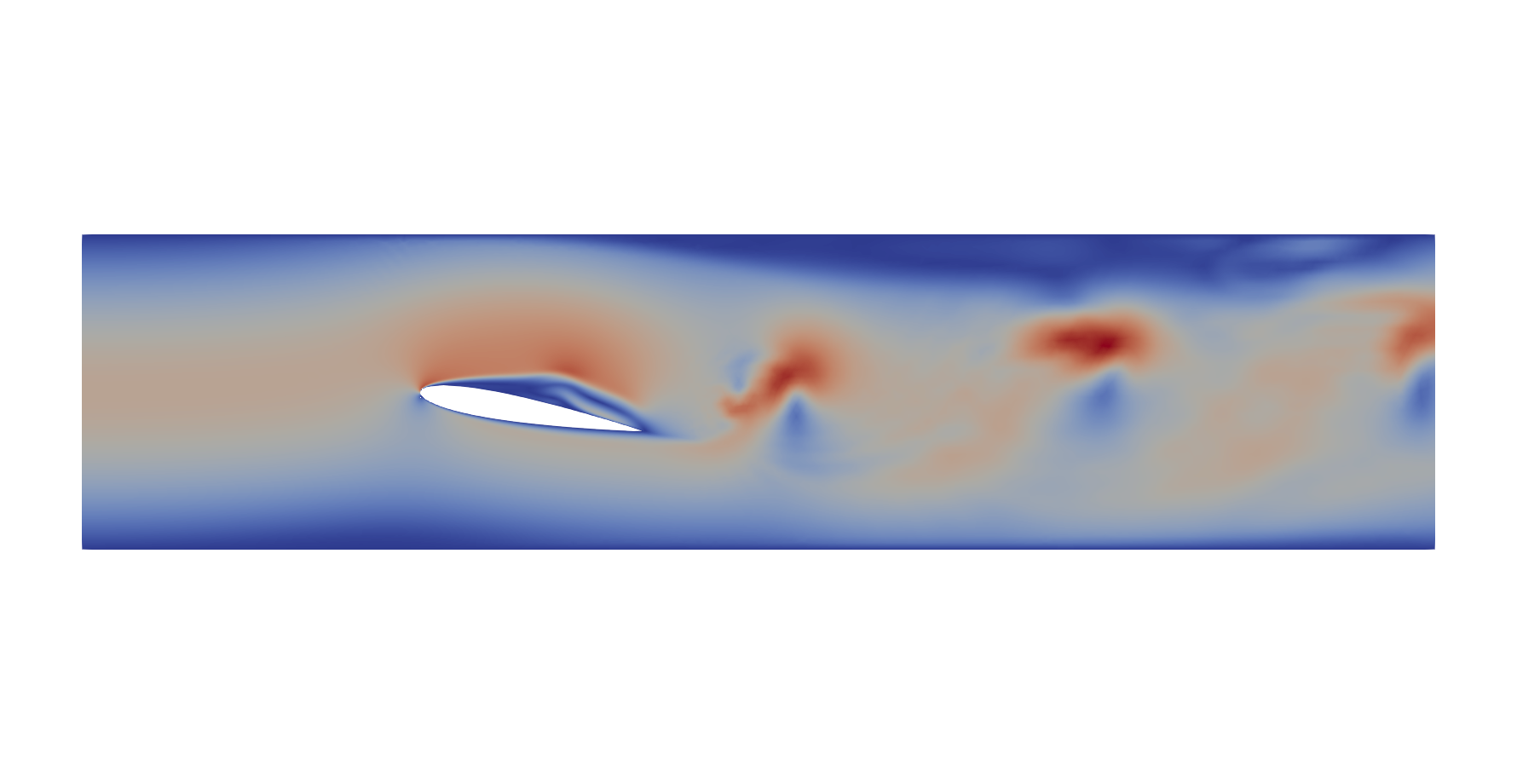}
    \caption{Velocity contour at the end of the baseline period before actuation begins. [$\sim$180 TU]}
    \label{fig:start}
\end{subfigure}
\vfill
\begin{subfigure}[b]{1\linewidth}
    \centering
    \includegraphics[trim={0 9cm 0 9cm},clip,width=0.7\linewidth]{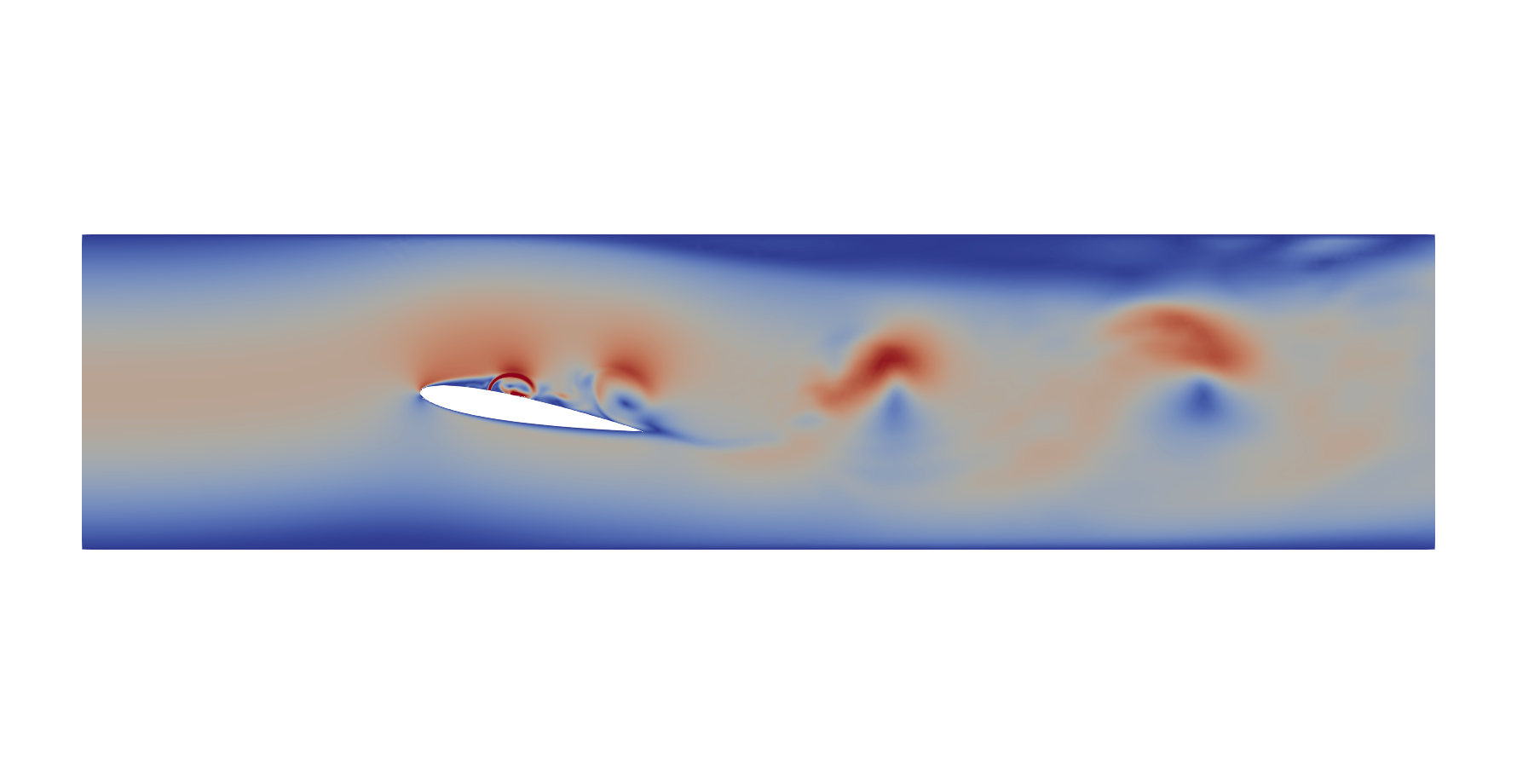}
    \caption{Velocity contour at the onset of the actuation phase. [$\sim$201 TU]}
    \label{fig:actuation}
\end{subfigure}
\vfill
\begin{subfigure}[b]{1\linewidth}
    \centering
    \includegraphics[trim={0 9cm 0 9cm},clip,width=0.7\linewidth]{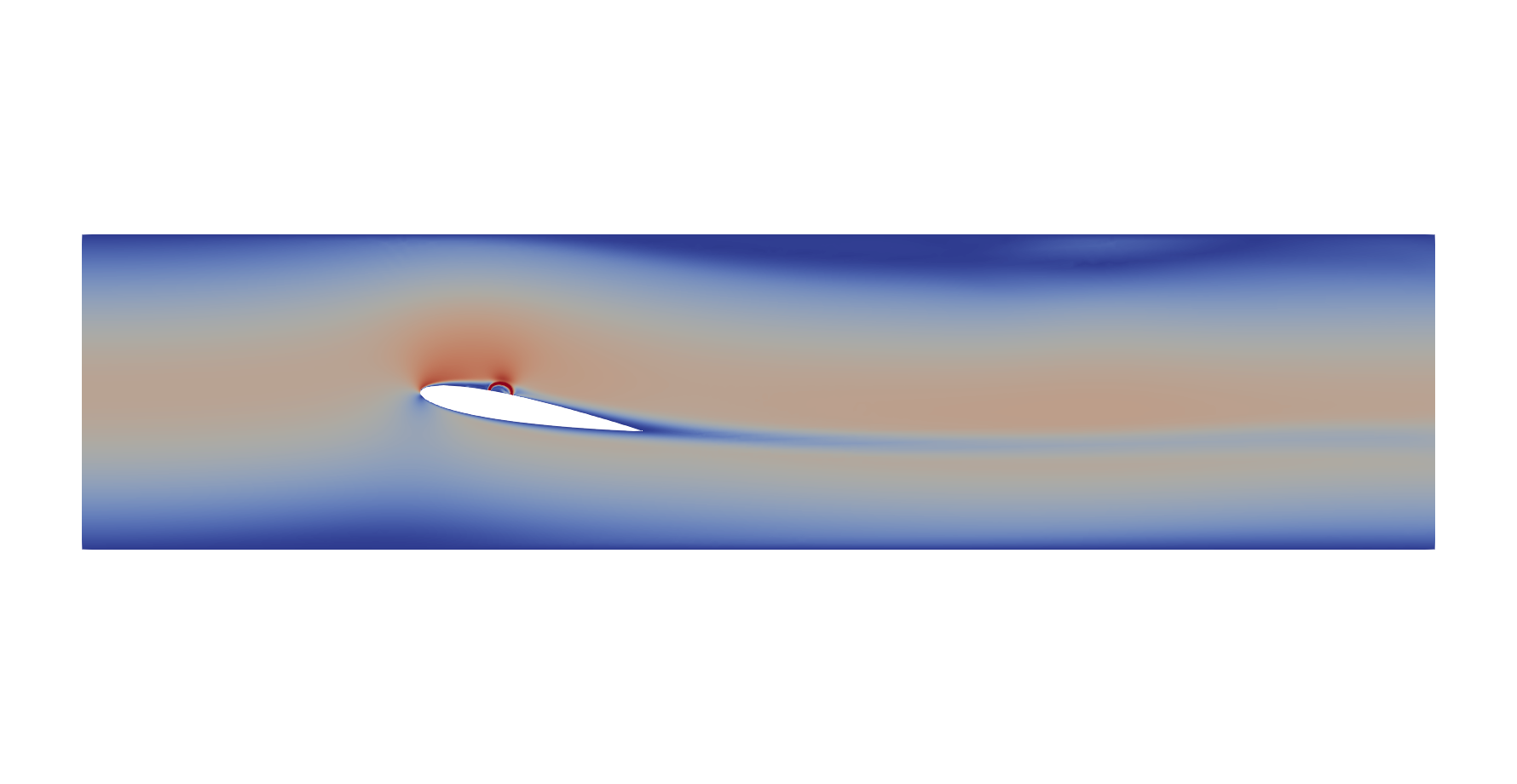}
    \caption{Velocity contour at the end of the actuation phase, illustrating a stabilized flow. [$\sim$240 TU]}
    \label{fig:stabilized}
\end{subfigure}

\raisebox{0cm}{\hspace{-1cm}%
    \begin{subfigure}[b]{1.25\linewidth}
        \centering
        \includegraphics[trim={0 7.5cm 0 7.5cm}, clip, width=1\linewidth]{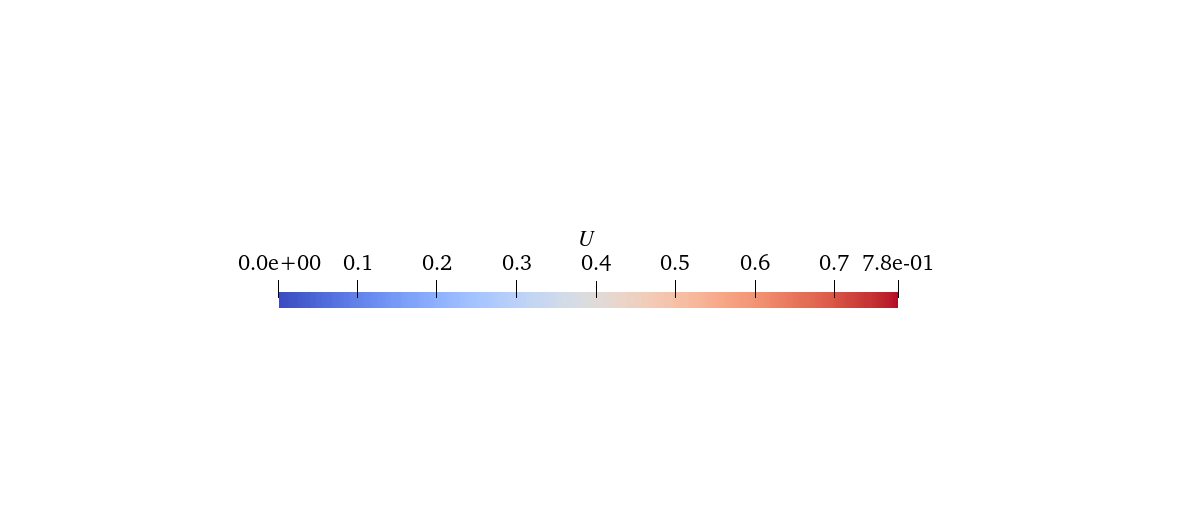}
    \end{subfigure}%
}

\caption{Comparison of flow conditions between baseline, onset of actuation, and stabilized flow using the fixed-intensity policy strategy from \citet{wang_deep_2022}.}
\label{fig:comparisonshotsflowfixed_wang}
\end{figure}

The results of the simulation, after reaching convergence, are shown in Table \ref{tab:results_1_fixed}. The evolution of the flow behavior and the aerodynamic coefficients can be observed in Figures \ref{fig:fixed_intensitiy_wang2} and \ref{fig:comparisonshotsflowfixed_wang}  A comparison between actuated and non-actuated flows reveals that vortex
 shedding generates a recirculation bubble near the trailing edge, causing localized separation and increased skin friction. The drag reduction achieved is approximately $28\%$, closely matching the $27\%$ reduction reported by \citet{wang_deep_2022}. However, a more pronounced difference is observed in the lift coefficient, with a $-3\%$ change in this study compared to a $+27\%$ increase in their results. It is likely that this discrepancy arises from the influence of ground effect, which limits further lift enhancement in this case. Despite this, aerodynamic efficiency improves by approximately $34\%$, indicating a significant gain even in the absence of lift augmentation.

An analysis of the flow dynamics reveals that once the jets are activated using a fixed-intensity control strategy, the flow stabilizes completely beyond \(115 \hspace{0.2cm} TU\), where $TU = tU_{\infty}/D$. Beyond this point, vortex shedding ceases to occur. The ability to achieve such flow stabilization and predictability using a simple and adaptable policy highlights the potential of this approach.

\begin{table}[htbp!]
\centering
\renewcommand{\arraystretch}{1.3}
\begin{tabular}{cccccc}
\hline \hline
\textbf{\(C_d\)} & \textbf{\(C_l\)} & \textbf{$C_l/C_d$} & \textbf{$\Delta C_l$} & \textbf{$\Delta C_d$} & \textbf{$\Delta C_l/C_d$} \\ \hline
$0.27$             & $1.77$             & $6.5$          & $-28.23\%$                    & -$2.89\%$                     & $+34.18\%$                        \\ \hline \hline
\end{tabular}
\caption{Results obtained using the jet intensities from Table \ref{tab:2} and comparison with baseline coefficients.}
\label{tab:results_1_fixed}
\end{table}

Once the strategy has been reproduced, an additional objective of this study is to explore alternative fixed-intensity control policies and evaluate their outcomes. To ensure a comprehensive analysis, a range of intensities from low to high has been examined. In total, 11 cases, including the reference case, have been simulated. These cases are summarized in Table \ref{tab:allcasesfixed}.

\begin{table}[htbp!]
\centering
\renewcommand{\arraystretch}{1.3}
\begin{tabular}{cccc}
\hline \hline
\textbf{Case} & \textbf{\(Q_{\mathrm{jet}_1}\)} & \textbf{\(Q_{\mathrm{jet}_2}\)} & \textbf{\(Q_{\mathrm{jet}_3}\)} \\ \hline
\#1  & -0.625        & 0.7           & -0.075        \\ 
\#2  & -0.8          & 0.9           & -0.1          \\ 
\#3  & -0.9          & 0.9           & -0.09         \\ 
\#4  & -1            & 1.1           & -0.1          \\ 
\#5  & -1.2          & 1.2           & -0.12         \\ 
\#6  & -1.25         & 1.25          & -0.125        \\ 
\#7  & -1.35         & 1.5           & -0.15         \\ 
\#8  & -1.5          & 1.35          & -0.15         \\ 
\#9  & -1.65         & 1.5           & -0.15         \\ 
\#10 & -1.8          & 2             & -0.2          \\ 
\#11 & -2.25         & 2.5           & -0.225        \\ \hline \hline
\end{tabular}
\caption{Definition of cases for the fixed-policy implementation, ranging from low to high mass flow rate intensities.}
\label{tab:allcasesfixed}
\end{table}

\begin{figure}[htbp!]
    \centering
    \includegraphics[trim={0cm 0cm 1cm 1.5cm},clip,width=0.8\linewidth]{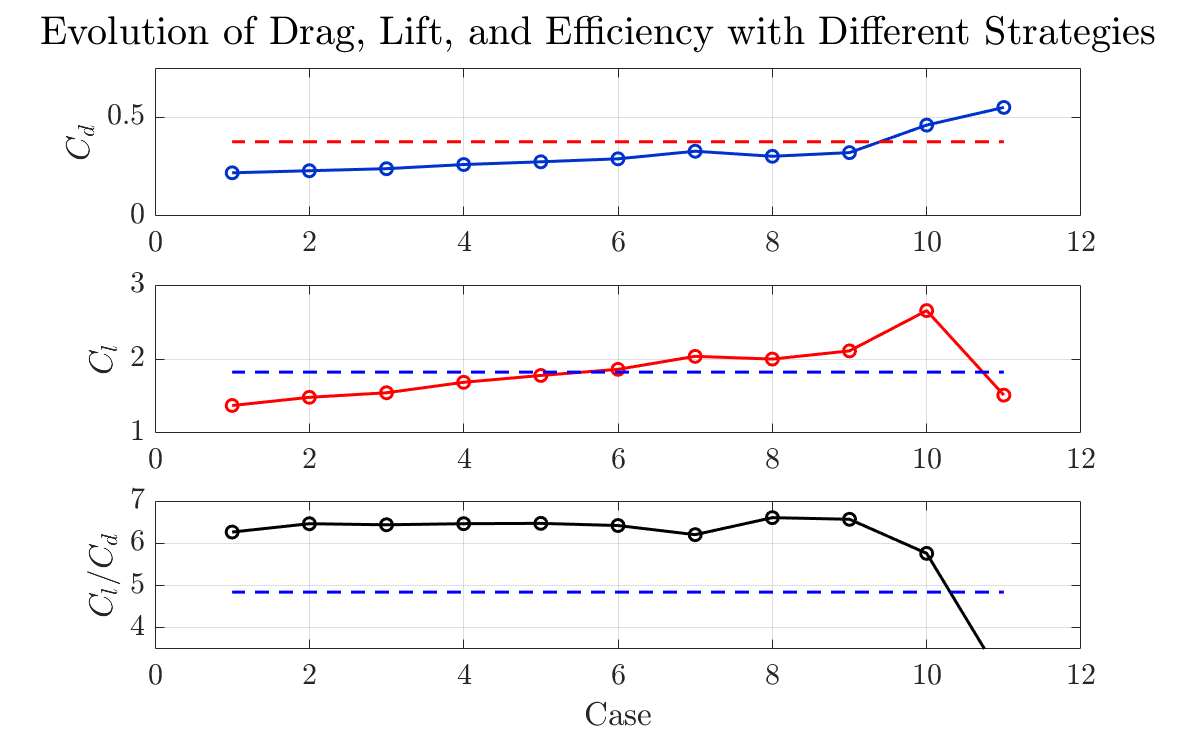}
    \caption{Aerodynamic coefficients resulting from the fixed-intensity policies listed in Table \ref{tab:allcasesfixed}, compared to the baseline values (represented by the dashed line).}
    \label{fig:allcasescomparisonfixed}
\end{figure}

The aerodynamic coefficients resulting from these implementations are shown in Figure \ref{fig:allcasescomparisonfixed}. The trends observed indicate that increasing the intensity while maintaining the same policy leads to an increase in both the drag coefficient \(C_d\) and the lift coefficient \(C_l\), while the aerodynamic efficiency remains relatively constant. This suggests that the selection of a particular control policy should depend on the specific optimization objective. For instance, if the primary goal is to reduce drag while maintaining lift, case \#5 would be a suitable choice. Conversely, if the objective is to maximize lift, case \#10 appears to be the optimal strategy.

A critical issue with excessive mass flow rate injection is the abrupt and unpredictable drop in aerodynamic performance. When the injected mass flow rate becomes too large, it disrupts the aerodynamic structures, leading to a loss of lift generation.

As shown in Figure \ref{fig:allcasescomparisonfixed}, the case achieving the greatest drag reduction is \#1, with a reduction of $38.1\%$. The highest increase in lift is observed in case \#10, with a $42.7\%$ increase. The most efficient overall case is \#8, yielding a $40.1\%$ improvement in aerodynamic efficiency.

\begin{figure}[htbp!]
\centering
\raisebox{0cm}{\hspace{0.5cm}%
    \begin{subfigure}[b]{0.9\linewidth}
        \centering
        \includegraphics[trim={0 8cm 0 8cm}, clip, width=1\linewidth]{colorbar_new.png}
    \end{subfigure}%
}
\begin{subfigure}[b]{1\linewidth}
    \centering
    \includegraphics[trim={0 0.5cm 0 0cm},clip,width=0.662\linewidth]{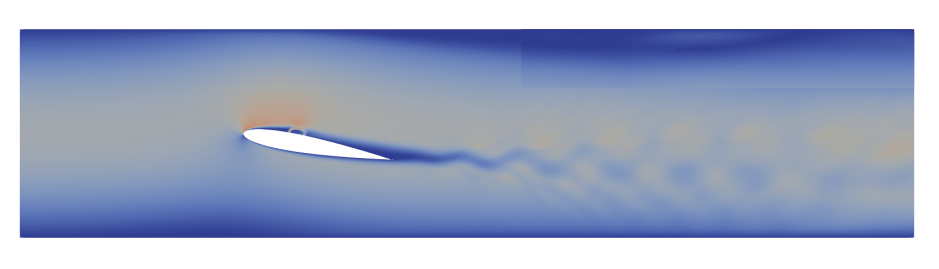}
    \caption{Low-intensity actuated flow.}
    \label{fig:low_intensity}
\end{subfigure}
\vfill
\begin{subfigure}[b]{1\linewidth}
    \centering
    \includegraphics[trim={0 9cm 0 9cm},clip,width=0.7\linewidth]{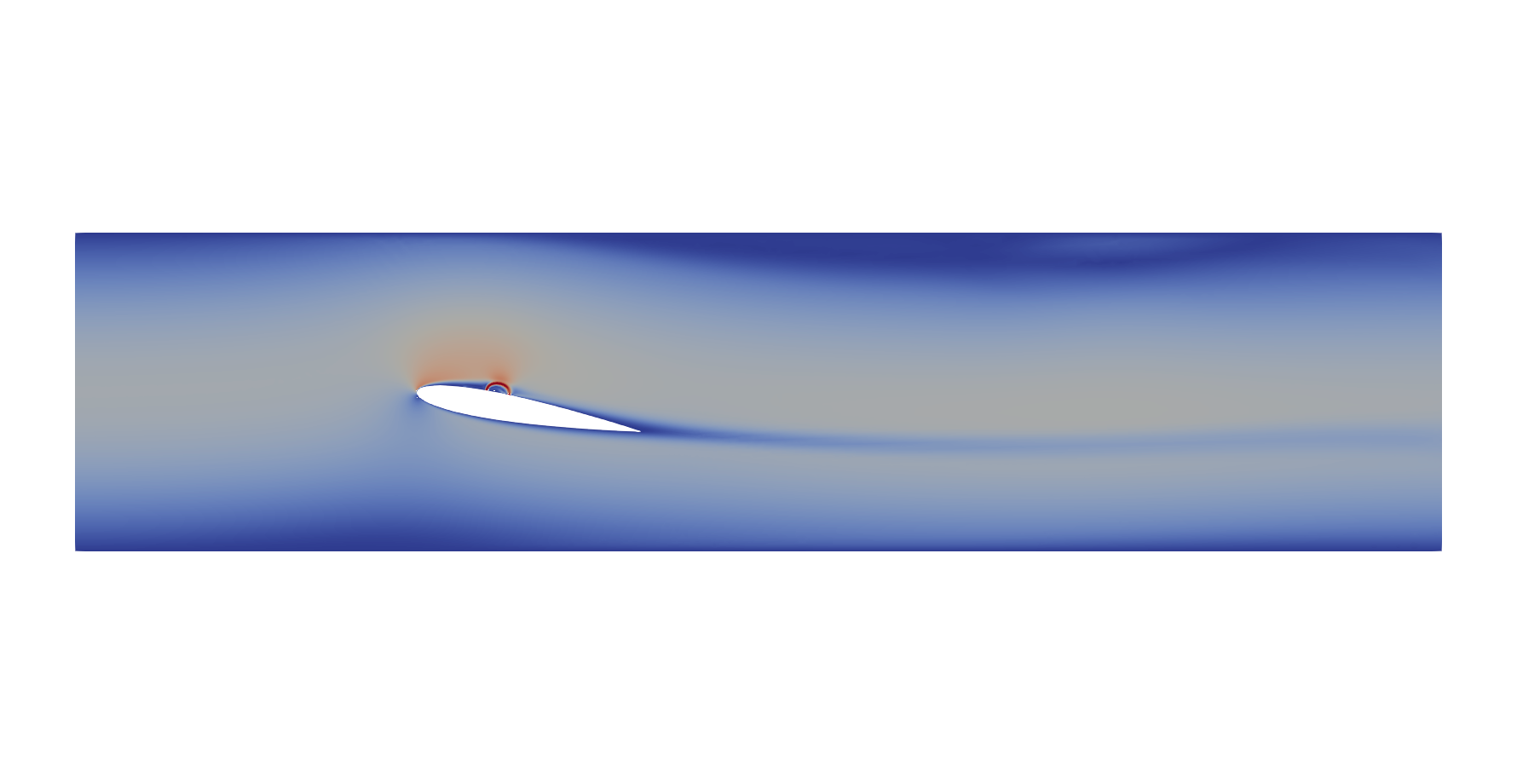}
    \caption{Medium-intensity actuated flow.}
    \label{fig:medium_intensity}
\end{subfigure}
\vfill
\begin{subfigure}[b]{1\linewidth}
    \centering
    \includegraphics[trim={0 9cm 0 9cm},clip,width=0.7\linewidth]{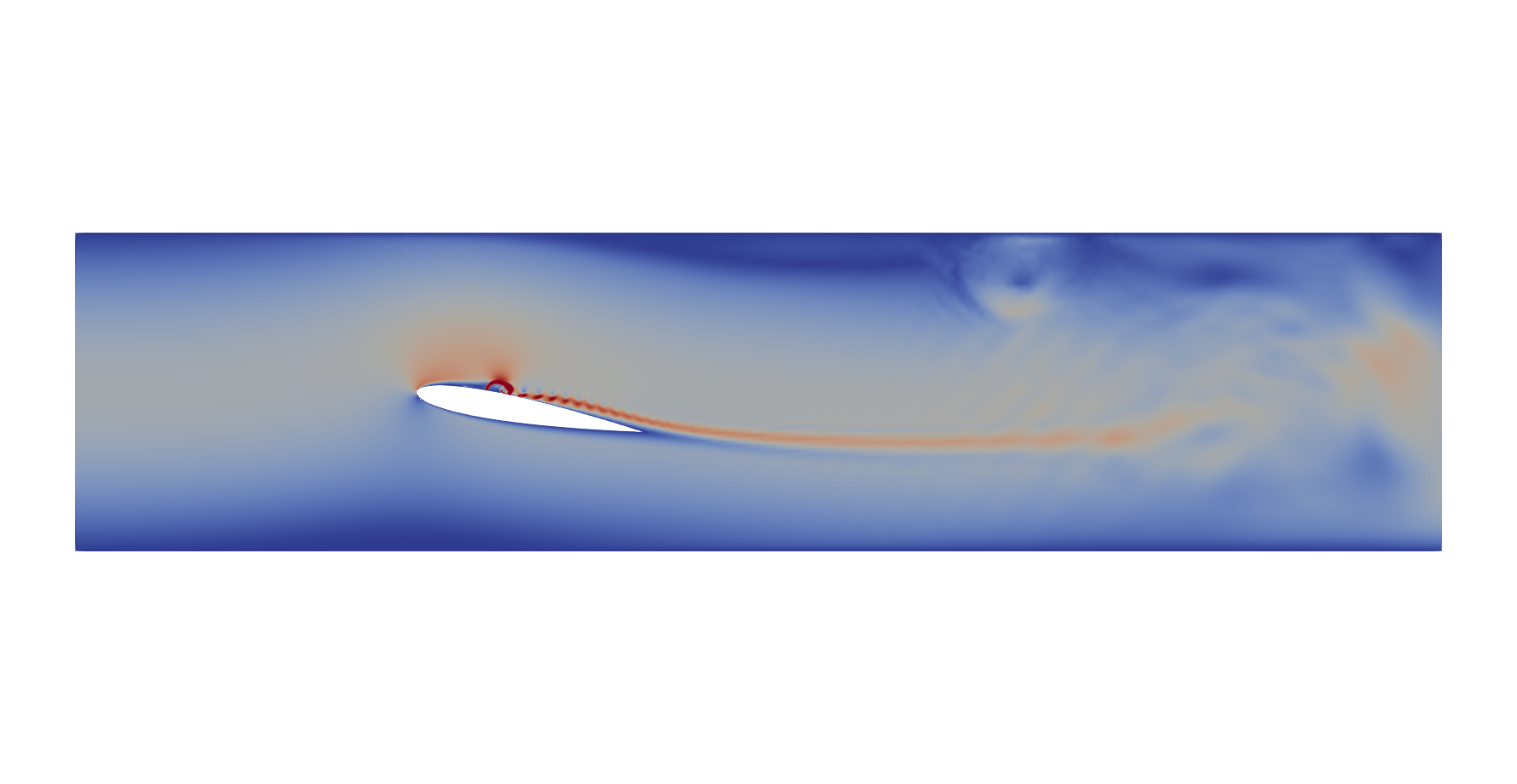}
    \caption{High-intensity actuated flow.}
    \label{fig:high_intensity}
\end{subfigure}
\caption{Comparison of flow behavior for low, medium, and high-intensity actuation cases using the fixed-intensity policy.}
\label{fig:comparisonshotsflowfixed}
\end{figure}

To further illustrate the flow behavior under different fixed-intensity actuation strategies, Figure \ref{fig:comparisonshotsflowfixed} presents a comparison of the stabilized flow fields for low, medium, and high intensity cases. In the low-intensity case, vortex shedding persists, though with a significantly reduced amplitude. In the medium-intensity case, the flow stabilizes completely, and no vortex shedding is observed. In the high-intensity case, the flow over the upper surface of the airfoil accelerates, leading to increased lift generation due to a higher pressure gradient.

\subsection{DRL training results of a 2D airfoil at \(Re_D\) \(3000\) using synthetic jets}

This section presents the main findings of the study, where the complete implementation of the framework is utilized to identify optimal strategies for flow control in two distinct directions. As previously explained, the objectives are to achieve drag reduction on one side and efficiency enhancement on the other.

\subsubsection{Drag reduction strategy case}\label{sec:drag_red}

\begin{figure}[htbp!]
    \centering
    \includegraphics[trim={0cm 0.5cm 1cm 0cm},clip,width=1\linewidth]{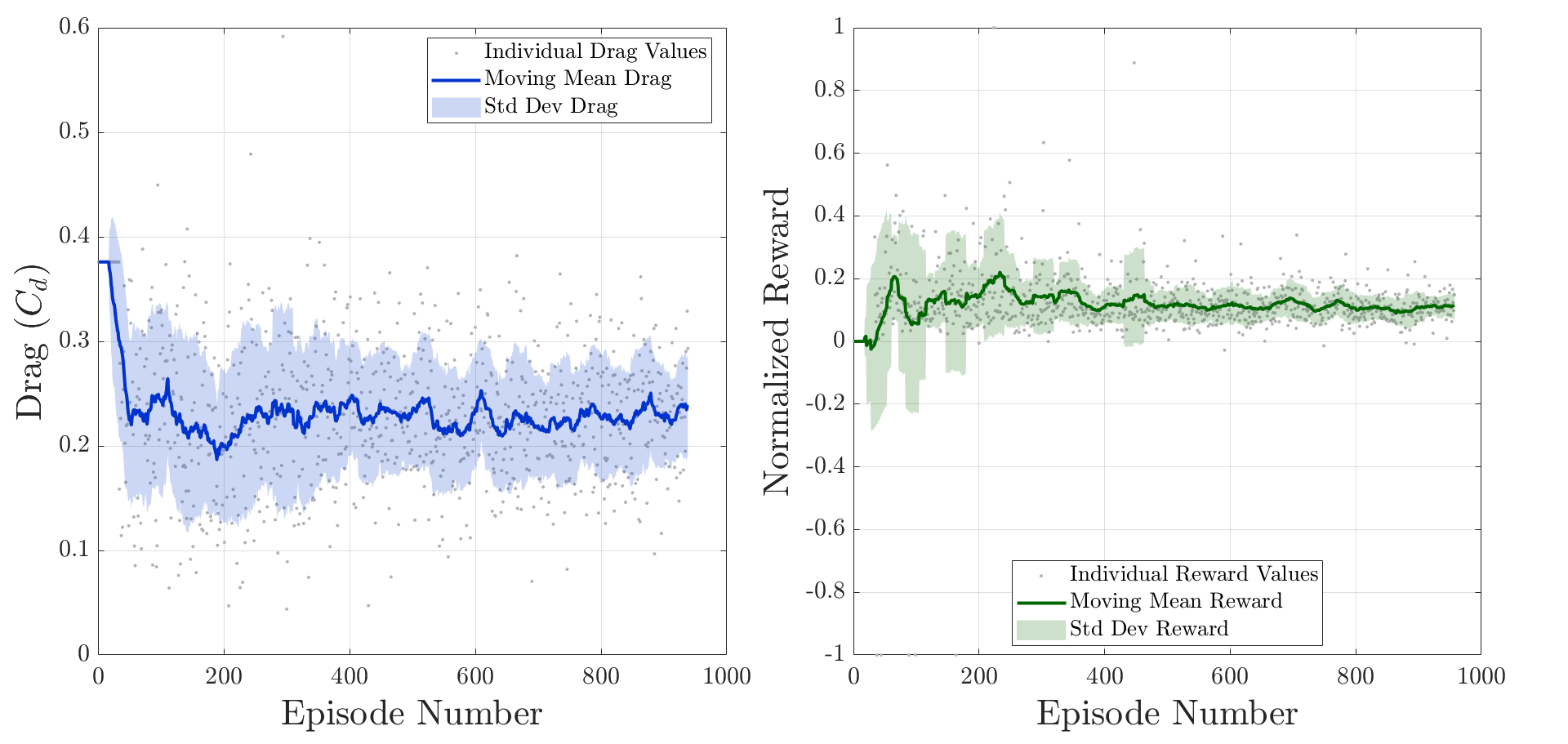}
    \caption{Comparison of the evolution of the drag coefficient ($C_d$) and normalized rewards during training. The left subplot illustrates drag values with a moving average and standard deviation, highlighting the trend and variability over episodes. The right subplot presents the rewards, showing progress and fluctuations in the reward signal. The shaded regions indicate the standard deviation for each metric.}
    \label{fig:train_evol_01}
\end{figure}

\begin{figure}[htbp!]
    \centering
    \begin{subfigure}[b]{0.49\linewidth}
        \centering
        \includegraphics[trim={0cm 0cm 0cm 0cm},clip,width=\linewidth]{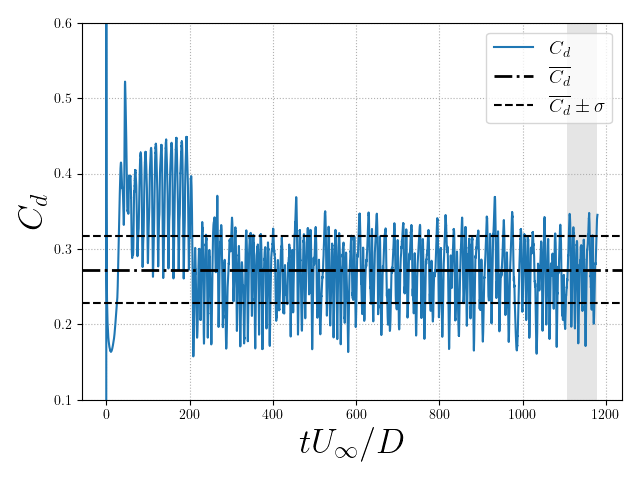}
        \caption{Drag coefficient throughout the deterministic case. Actuation starts at 200 TU.}
        \label{fig:drag01}
    \end{subfigure}
    \hfill
    \begin{subfigure}[b]{0.49\linewidth}
        \centering
        \includegraphics[trim={0cm 0cm 0cm 0cm},clip,width=\linewidth]{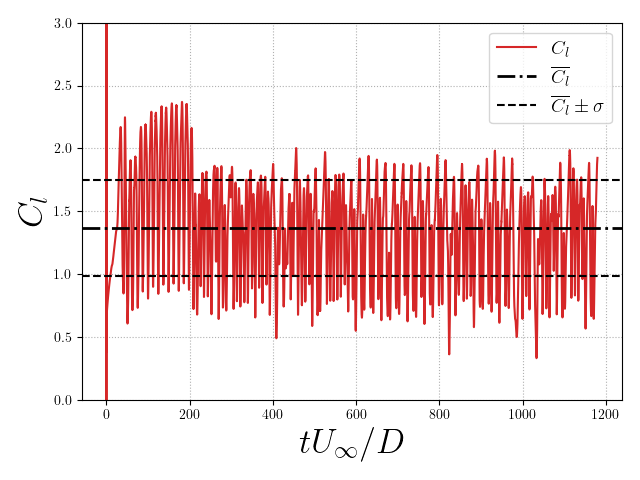}
        \caption{Lift coefficient throughout the deterministic case. Actuation starts at 200 TU.}
        \label{fig:lift01}
    \end{subfigure}
    \caption{Evolution of drag and lift coefficients throughout the deterministic case, with actuation commencing at 200 TU.}
    \label{fig:drag_lift_combined}
\end{figure}

In the first case, focusing on drag reduction, a total of 957 episodes were used to train the agent. The evolution of the training process is presented in Figure \ref{fig:train_evol_01}. The final deterministic case obtained from this training is depicted in Figures \ref{fig:drag01} and \ref{fig:lift01}. 

\begin{figure}[htbp!]
    \centering
    \begin{subfigure}[b]{0.49\linewidth}
        \centering
        \includegraphics[trim={0cm 0cm 0cm 0cm},clip,width=\linewidth]{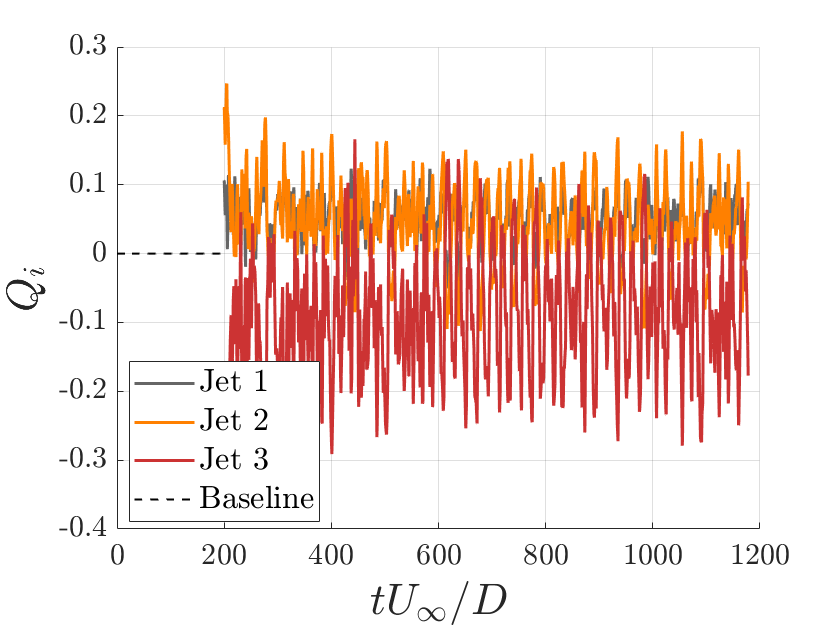}
        \caption{Jet actions during the deterministic case for drag reduction.}
        \label{fig:actions_01}
    \end{subfigure}
    \hfill
    \begin{subfigure}[b]{0.49\linewidth}
        \centering
        \includegraphics[trim={0cm 0cm 0cm 0cm},clip,width=\linewidth]{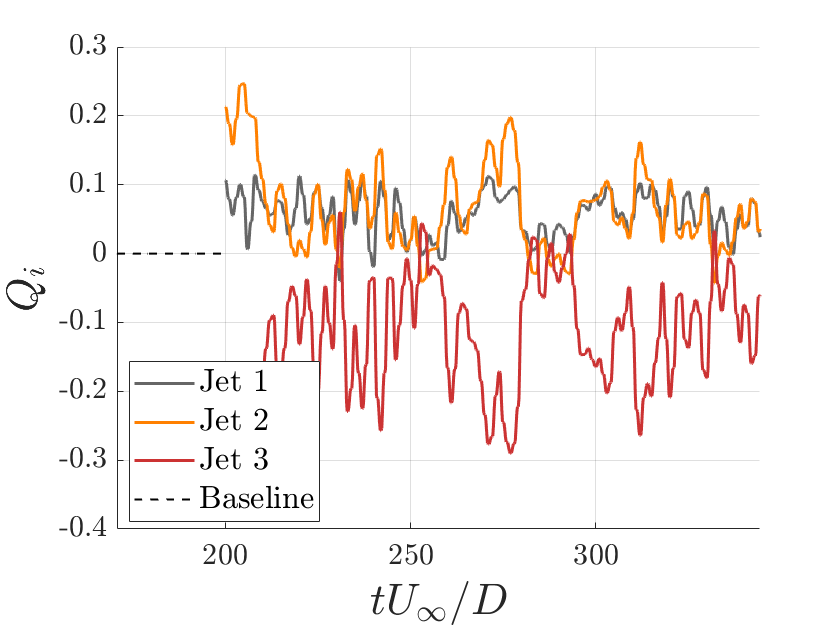}
        \caption{Detailed view of jet actions over a specific time range, providing insight into the agent's control decisions.}
        \label{fig:actions_01_zoom}
    \end{subfigure}
    \caption{Comparison of jet actions during the deterministic case for drag reduction, including a detailed analysis of a specific time range.}
    \label{fig:actions_combined}
\end{figure}

By implementing the control actions illustrated in Figure \ref{fig:actions_01}, with further details provided in Figure \ref{fig:actions_01_zoom}, the results indicate an average drag reduction of 43.89\% and a lift reduction of 35.89\% during the actuated phase. Compared to the best drag reduction case from Section \ref{sec:intensityfixed}, the DRL-based strategy achieves a 6\% greater drag reduction. 

The control actions employed by the agent exhibit a distinct pattern rather than remaining constant. The jets alternate between blowing (adding momentum) and suction (reducing momentum) throughout the control process. Jets 1 and 2 exhibit similar actuation patterns, indicating that they perform nearly identical actions. In contrast, Jet 3 consistently executes the opposite action to Jets 1 and 2, ensuring that the net mass flow rate remains zero.

\begin{figure}[htbp!]
    \centering
    \includegraphics[trim={6cm 5cm 6cm 0cm},clip,width=0.8\linewidth]{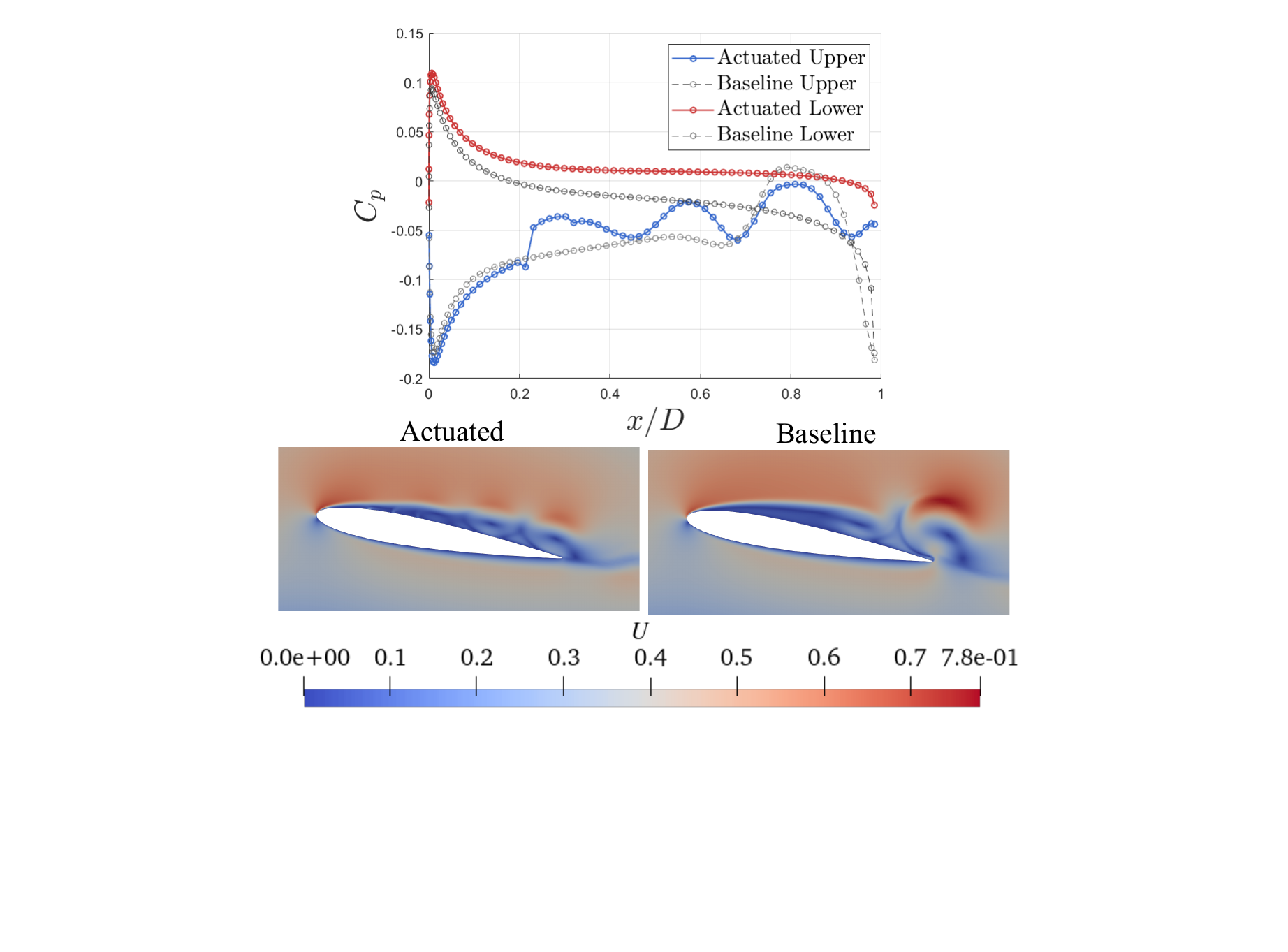}
    \caption{Comparison of \( C_p \) along the airfoil for the baseline (right) and actuated (left) cases. Velocity contours for the instants used to extract the \( C_p \) distribution are also shown.}
    \label{fig:cp_01}
\end{figure}

This result can be further analyzed by examining Figure \ref{fig:cp_01}, which presents the pressure coefficient distribution along the airfoil for both the actuated drag reduction case and the baseline. The figure illustrates how the jets decelerate the flow over the upper surface, leading to a reduction in lift. Additionally, it is important to highlight that the small oscillations near the trailing edge diminish in magnitude and smoothly converge with the lower surface, thereby forming a weaker wake. This process effectively reduces the primary source of drag.

\begin{figure}[htbp!]
    \centering
    \begin{subfigure}[b]{0.49\linewidth}
        \centering
        \raisebox{0cm}{ 
            \includegraphics[trim={0cm 0cm 0cm 0cm},clip,width=\linewidth]{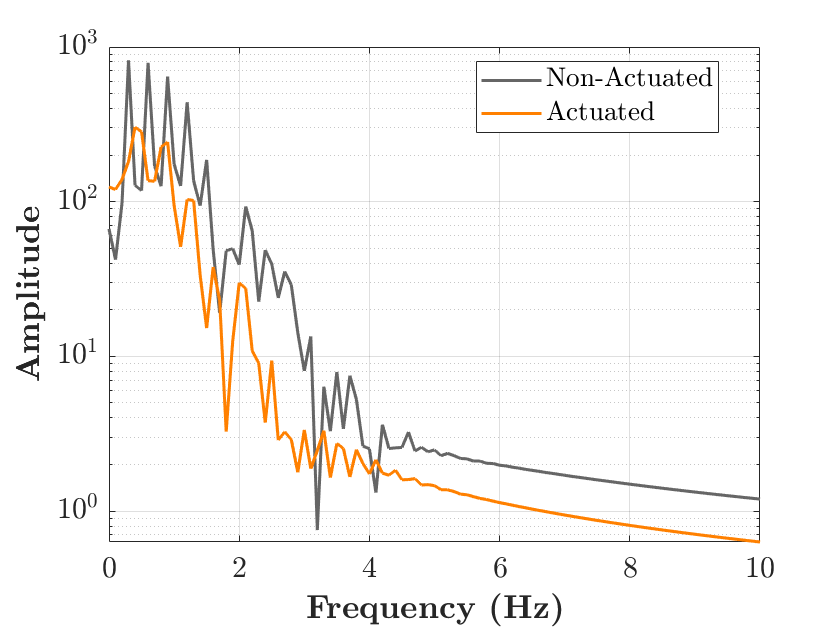}
        }
        \caption{Frequency spectrum comparison of the wake for non-actuated and actuated cases. Actuation significantly reduces the amplitude of dominant frequencies, leading to a weakened wake and reduced vortex shedding.}
        \label{fig:fft01}
    \end{subfigure}
    \hfill
    \begin{subfigure}[b]{0.49\linewidth}
        \centering
        \raisebox{0.2cm}{ 
            \includegraphics[trim={0cm 0cm 0cm 0cm},clip,width=\linewidth]{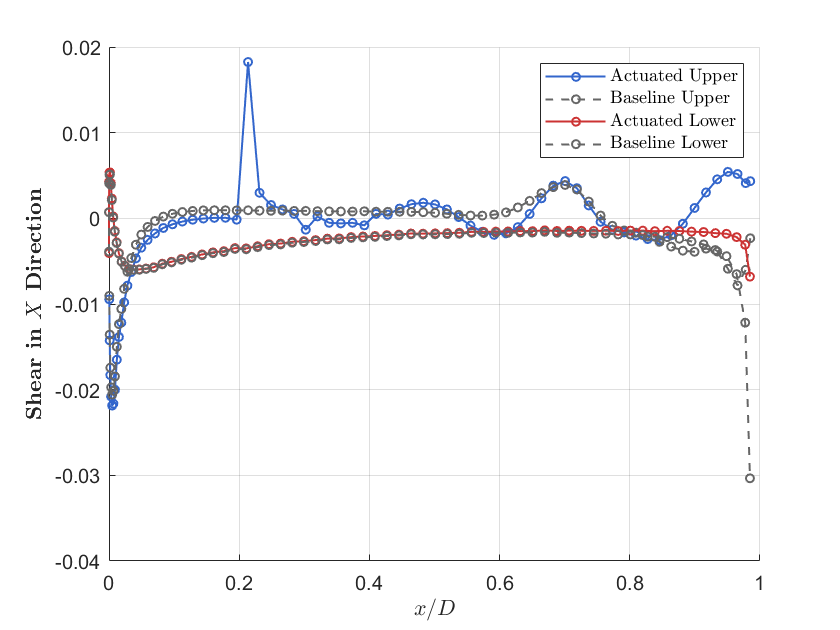}
        }
        \caption{Comparison of streamwise shear stress along the airfoil surface for actuated and baseline cases, highlighting the impact of actuation on boundary layer behavior and attachment.}
        \label{fig:shear01}
    \end{subfigure}
    \caption{Comparison of frequency spectrum (left) and streamwise shear stress (right) for non-actuated and actuated cases, demonstrating the effect of actuation on wake dynamics and boundary layer behavior.}
    \label{fig:fft_shear_combined}
\end{figure}

The primary mechanism responsible for drag reduction is the control of vortex shedding and the associated recirculation bubble. From the perspective of vortex shedding, Figure \ref{fig:fft01} presents the pressure signal recorded at a witness point located at \([x/D,y/D] = [2,0]\) in the wake region. The results indicate that the jet-induced strategy reduces the amplitude of oscillatory shedding. The magnitude of these oscillations is decreased by approximately 62.5\%, allowing the dissipation of discontinuities at a faster rate and preventing the formation of strong eddies in the wake, which are otherwise observed in the non-actuated flow field.

Another major contributor to drag is the boundary layer and the resulting skin friction \citet{orlu_instantaneous_2020}. It is essential to assess whether the boundary layer remains attached or detaches and whether the jets exert sufficient control authority to influence its behavior. This is illustrated in Figure \ref{fig:shear01}, which shows the streamwise shear stresses around the airfoil. The figure highlights how the jets energize the boundary layer, with the first jet actively injecting momentum. Beyond the jet region, the oscillatory patterns seen in the \( C_p \) plot are also visible in the shear stress distribution, indicating multiple instances of detachment and reattachment before reaching the trailing edge. A comparison between actuated and non-actuated flows reveals that vortex shedding generates a recirculation bubble near the trailing edge, causing localized separation and increased skin friction. The clearest visualization of the recirculation bubble is provided in Figure \ref{fig:cp_01}, where the non-actuated velocity field is depicted.

\begin{figure}[htbp!]
    \centering
    \begin{subfigure}[b]{0.8\linewidth}
        \centering
        \includegraphics[trim={0cm 0cm 1cm 0cm},clip,width=\linewidth]{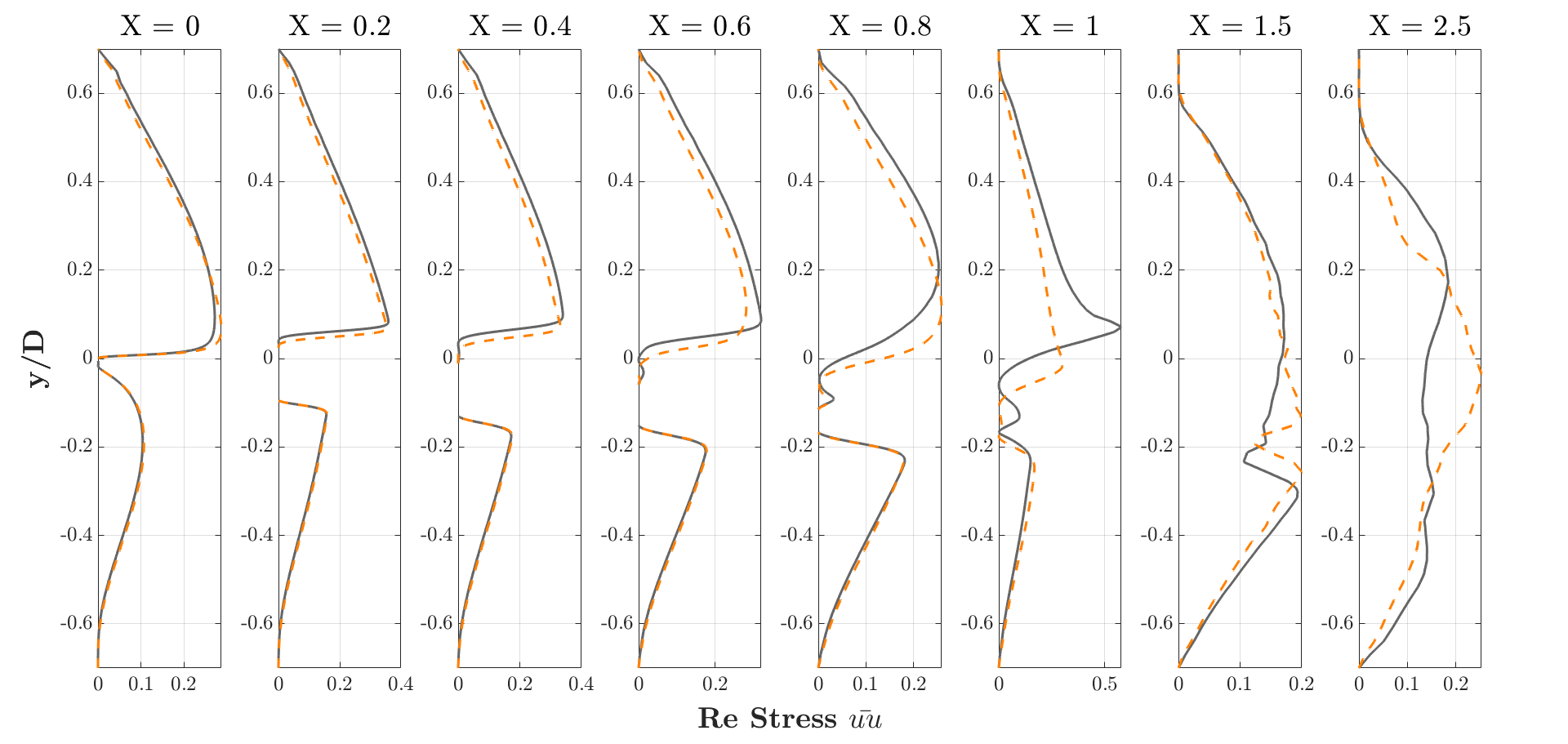}
        \caption{Reynolds stress \( \bar{uu} \).}
    \end{subfigure}
    \vfill
    \begin{subfigure}[b]{0.8\linewidth}
        \centering
        \includegraphics[trim={0cm 0cm 1cm 0cm},clip,width=\linewidth]{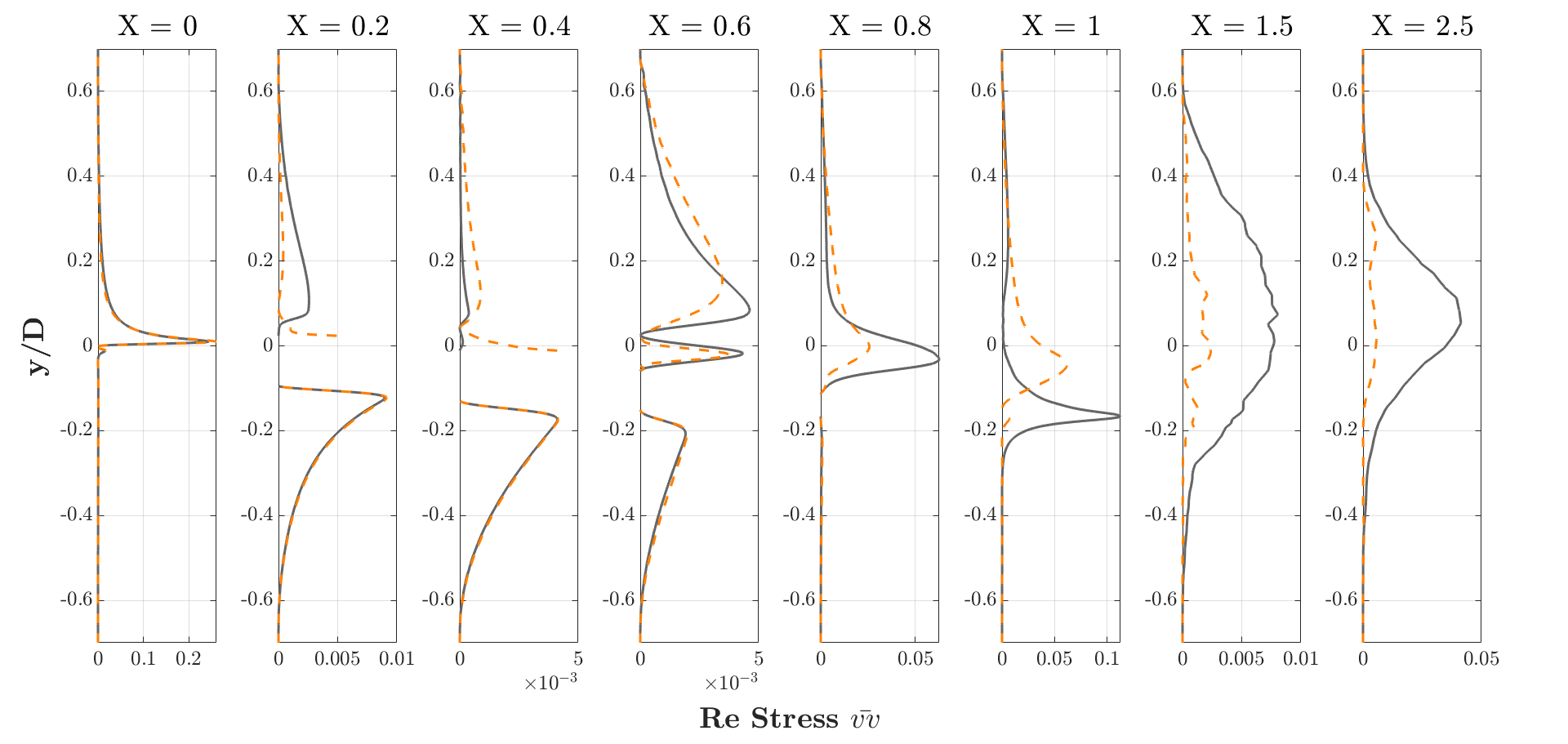}
        \caption{Reynolds stress \( \bar{uv} \).}
    \end{subfigure}
    \vfill
    \begin{subfigure}[b]{0.8\linewidth}
        \centering
        \includegraphics[trim={0cm 0cm 1cm 0cm},clip,width=\linewidth]{re_vv.png}
        \caption{Reynolds shear stress \( \bar{vv} \).}
    \end{subfigure}
    \caption{Comparison of Reynolds stresses (\( \bar{uu} \), \( \bar{uv} \), and \( \bar{vv} \)) along the \( Y \)-coordinate at different streamwise locations (\( x/D \)). The dashed orange lines represent the actuated case, while the solid grey lines correspond to the baseline flow.}
    \label{fig:reynolds_stresses_combined}
\end{figure}

By analyzing the Reynolds stresses at different streamwise locations, valuable insights into the turbulent structures and flow characteristics can be obtained. Figure \ref{fig:reynolds_stresses_combined}a presents the Reynolds stresses along the streamwise direction, where the oscillations in both the baseline and actuated flows exhibit noticeably smaller magnitudes. At \( x/D = 2.5 \), in the middle of the wake, the expected streamwise turbulence is observed, as the inflow follows a parabolic profile. This suggests that the actuated wake more closely resembles the inflow compared to the baseline case, reinforcing the drag reduction mechanism through improved pressure distribution. A similar trend is observed in Figures \ref{fig:reynolds_stresses_combined}b and \ref{fig:reynolds_stresses_combined}c, where wake positions exhibit lower stress values. When turbulence-related stresses are captured, they are primarily concentrated in the middle of the wake, as expected.

In Figure \ref{fig:reynolds_stresses_combined}b, the effects of jet actuation are evident, leading to significant turbulence mixing in the wall-normal direction. However, this influence results in a smaller turbulent region along the upper surface of the airfoil compared to the non-actuated case. While turbulence mixing persists, its intensity is reduced to half the magnitude observed in the non-actuated flow, with this reduction becoming even more pronounced when analyzing turbulence in the wake region.

Figure \ref{fig:reynolds_stresses_combined}c exhibits a similar behavior along the airfoil. The first notable discrepancies arise around \( x/D = 1 \), indicating that the non-actuated flow field experiences interaction between the streamwise and wall-normal directions due to the recirculation bubble formed in that region on both the upper and lower surfaces. In contrast, for the actuated case, turbulence mixing and eddy generation are restricted to a single peak, suggesting the absence of a recirculation bubble, as the flow field predominantly mixes in a single direction. Beyond this region, in the wake, the interaction between turbulent eddies and the main stream flow decays significantly in the actuated case compared to the non-actuated flow, indicating a weaker and more stable wake.

\subsubsection{Efficiency enhancement strategy case}

For the second case, different training strategies have been employed to optimize the agent's performance. As described in Section \ref{sec:DRL_characteristics}, two distinct reward formulations have been utilized. In the presented deterministic case, the agent is trained using the third reward function, which separates lift and drag while assigning different weights to each parameter. Specifically, weights of $0.3$ and $0.7$ were applied to drag and lift, respectively, following Equation \ref{eq:reward_393}.

\begin{figure}[htbp!]
    \centering
    \includegraphics[trim={0.2cm 0.5cm 1cm 0cm},clip,width=1\linewidth]{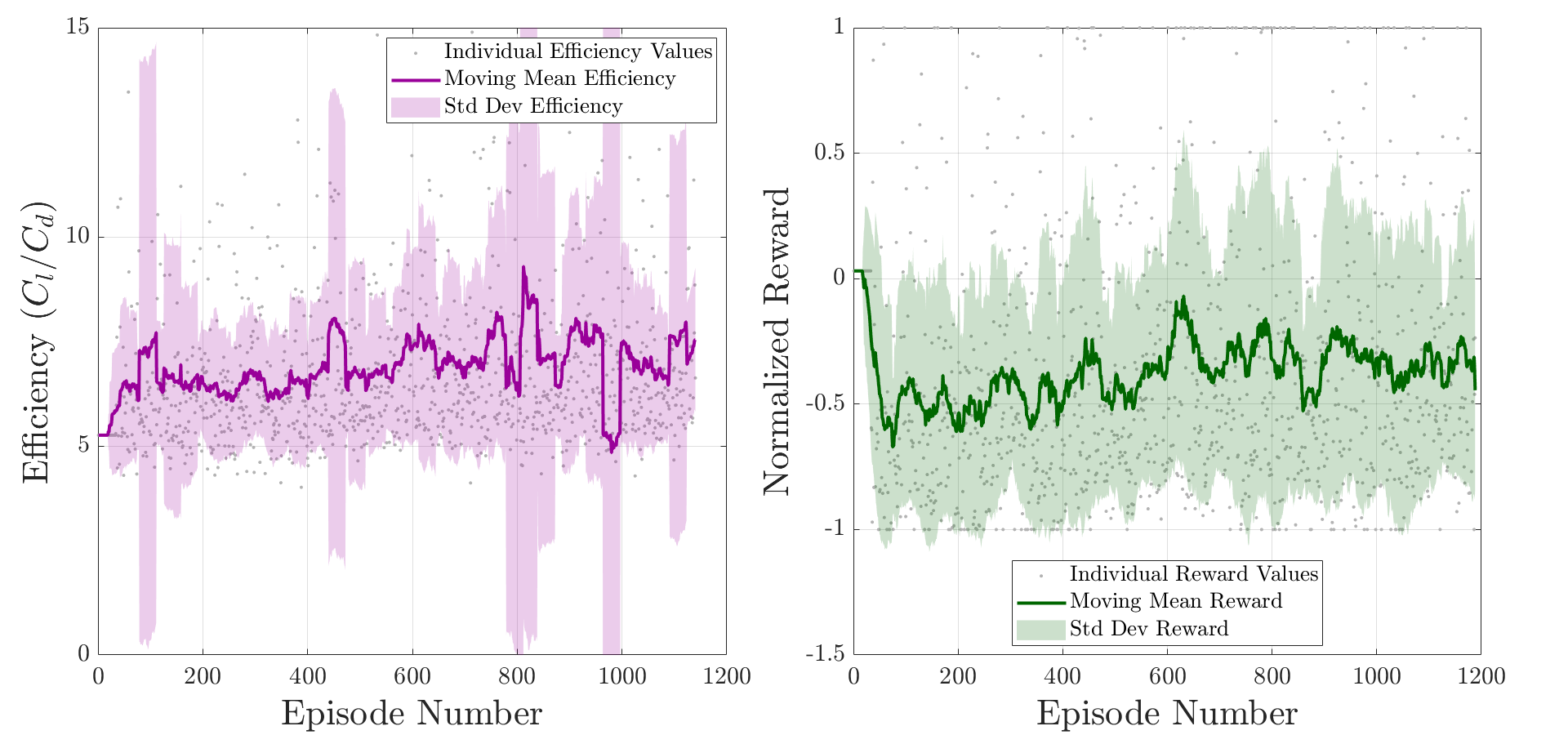}
    \caption{Comparison of the evolution of aerodynamic efficiency ($C_l/C_d$) and normalized rewards during training. The left subplot illustrates efficiency values with a moving average and standard deviation, capturing trends and variability over episodes. The right subplot presents the reward evolution, highlighting progress and fluctuations. The shaded regions indicate standard deviation for each metric.}
    \label{fig:train_evol_B2}
\end{figure}

A total of 1188 episodes were required for training, with the evolution of the training process illustrated in Figure \ref{fig:train_evol_B2}.

\begin{figure}[htbp!]
    \centering
    \begin{subfigure}[b]{0.48\linewidth}
        \centering
        \includegraphics[trim={0cm 0cm 0cm 0cm},clip,width=\linewidth]{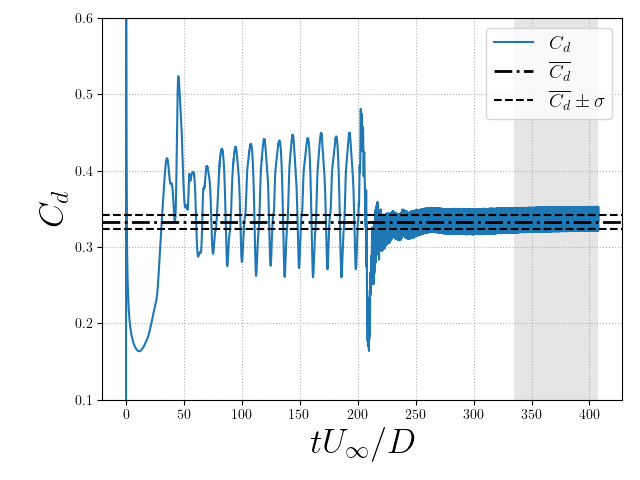}
        \caption{Drag coefficient ($C_d$) evolution over time, showing oscillatory behavior during the transient phase and a reduction in drag in the steady-state region under DRL-based active flow control.}
        \label{fig:dragB2}
    \end{subfigure}
    \hfill
    \begin{subfigure}[b]{0.48\linewidth}
        \centering
        \includegraphics[trim={0cm 0cm 0cm 0cm},clip,width=\linewidth]{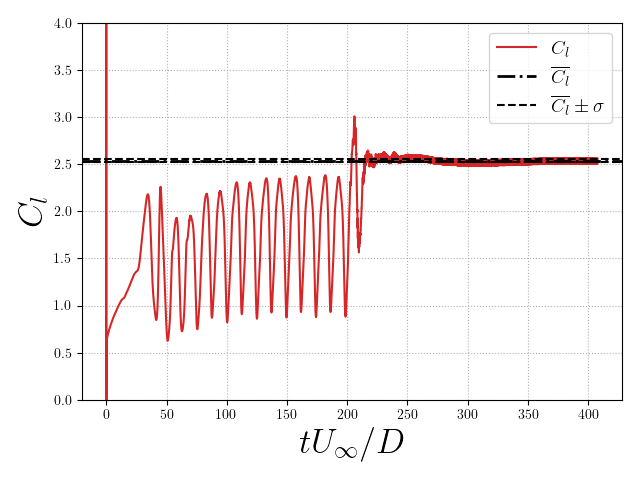}
        \caption{Lift coefficient ($C_l$) evolution over time, illustrating initial transient oscillations and stabilization at a higher value due to jet actuation. The shaded region represents steady-state statistics.}
        \label{fig:liftB2}
    \end{subfigure}
    \caption{Time evolution of aerodynamic coefficients under DRL-based active flow control: (a) Drag coefficient ($C_d$), showing steady-state reduction; (b) Lift coefficient ($C_l$), highlighting stabilization at higher values.}
    \label{fig:cd_cl_combined}
\end{figure}

\begin{figure}[htbp!]
    \centering
    \includegraphics[trim={0cm 0cm 0cm 0cm},clip,width=0.5\linewidth]{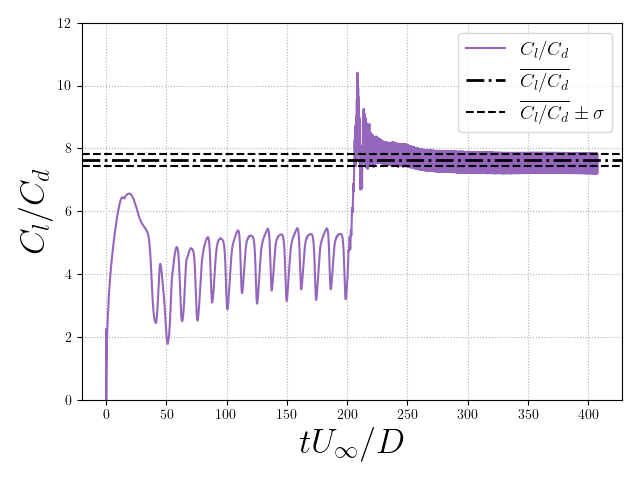}
    \caption{Aerodynamic efficiency ($C_l/C_d$) over time, illustrating the improvement achieved through DRL-based active flow control. The shaded region represents time-averaged steady-state values with standard deviation.}
    \label{fig:effB2}
\end{figure}

The results of the deterministic strategy are presented in Figures \ref{fig:dragB2}, \ref{fig:liftB2}, and \ref{fig:effB2}, depicting the evolution of drag, lift, and aerodynamic efficiency, respectively. The final outcome yields an 11.97\% reduction in drag, a 39.63\% increase in lift, and an overall efficiency improvement of 58.64\%. When compared to the best fixed-intensity case from Section \ref{sec:intensityfixed}, the DRL agent achieves an additional 18\% improvement in aerodynamic efficiency.

\begin{figure}[htbp!]
    \centering
    \includegraphics[trim={0cm 0cm 0cm 0cm},clip,width=0.5\linewidth]{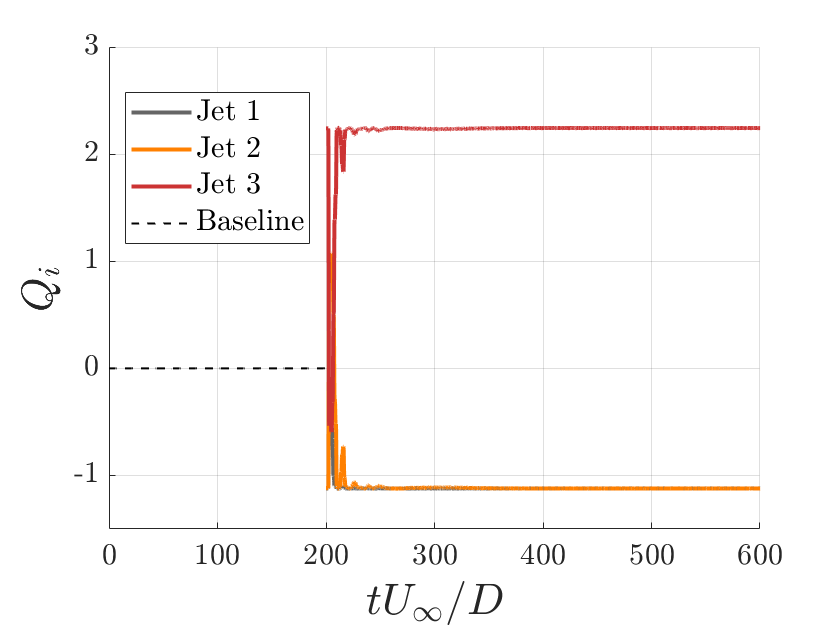}
    \caption{Time evolution of the mass flow rate \( Q \) for the three jets, illustrating the transition from the baseline to the steady-state actuation phase.}
    \label{fig:actionsB2}
\end{figure}

As observed in Figure \ref{fig:actionsB2}, the control actions remain nearly constant throughout the actuation phase. This trend has been consistently identified in all efficiency-driven cases. Regardless of the specific reward function or parameter adjustments, the agent converges toward a steady-state strategy, with a transition phase lasting no longer than 15 to 20 actions.

\begin{figure}[htbp!]
    \centering
    \includegraphics[trim={6cm 5cm 6cm 0cm},clip,width=0.8\linewidth]{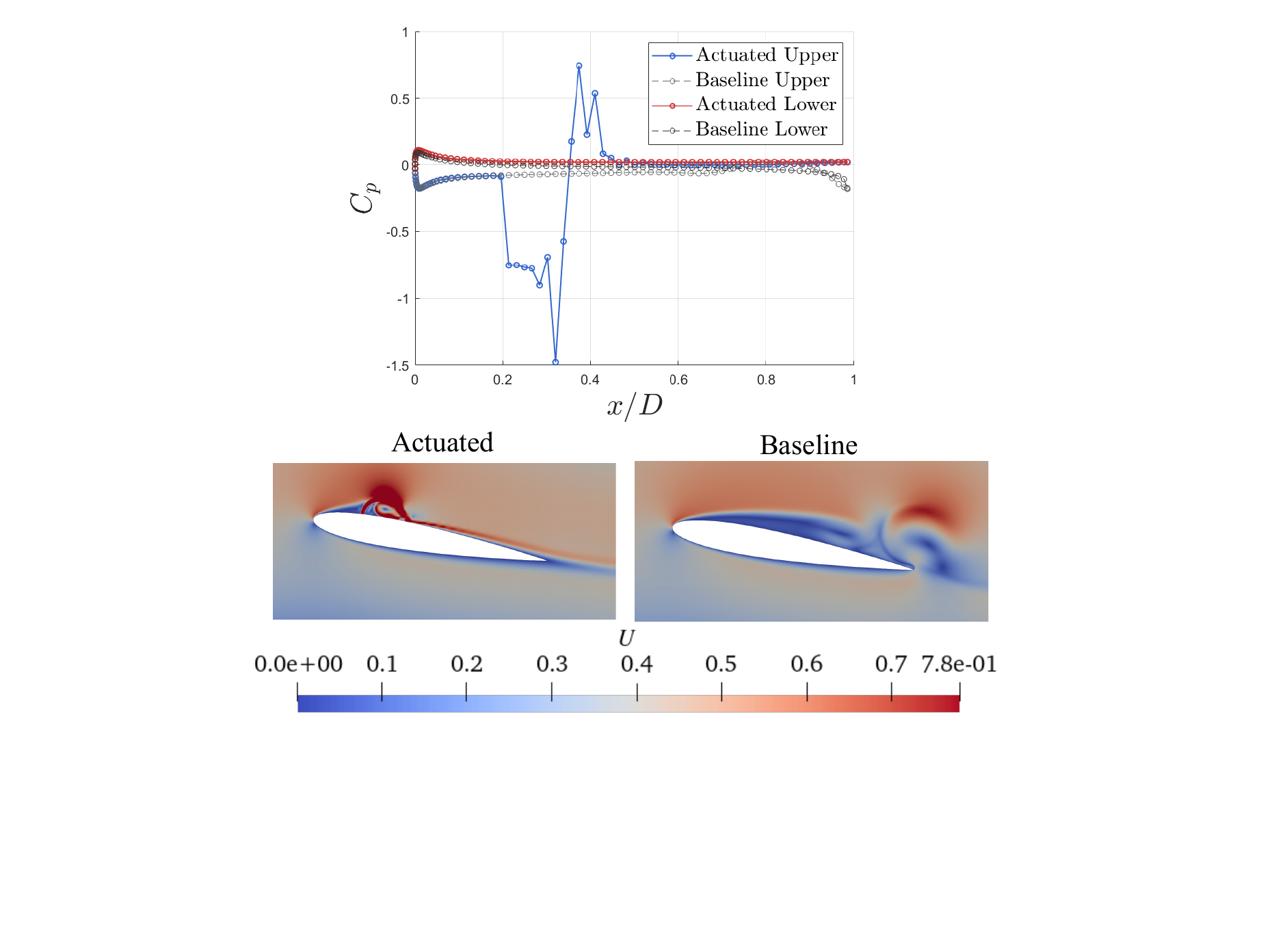}
    \caption{Comparison of \( C_p \) along the airfoil for the baseline (right) and actuated (left) cases. Velocity contours are displayed for the instants used to extract the \( C_p \) distribution.}
    \label{fig:cp_B2}
\end{figure}

Notably, the control actions in this case are approximately ten times more aggressive than those used in the drag reduction case discussed in Section \ref{sec:drag_red}. By employing a higher mass flow rate, the agent and jets effectively transform the baseline dynamic state, characterized by vortex shedding and a recirculation bubble, into a completely steady-state flow, controlled by the induced jet structure. As illustrated in Figures \ref{fig:actionsB2} and \ref{fig:cp_B2}, $\mathrm{jet}_1$ and $\mathrm{jet}_2$ primarily function to counteract the mass flow introduced by $\mathrm{jet}_3$. This interaction is particularly significant and warrants deeper investigation.

The resulting flow structure closely resembles that identified in \citet{wang_deep_2022}, though with a key difference: in their study, jet 2 was responsible for adding mass flow, while jet 1 facilitated the formation of a circular flow structure. In contrast, in the present study, the generated recirculation bubble slightly modifies the upper region of this structure, leading to an elliptical rather than a circular shape. This elliptical flow pattern accelerates in the direction of the airfoil before realigning with the main flow. 

A detailed examination of Figure \ref{fig:cp_B2} reveals that, prior to jet actuation, the region of highest flow acceleration creates a near-zero velocity zone, effectively forming a void. The jets mitigate this effect by generating a continuous flow structure that not only accelerates the flow in the initial region but also decelerates it downstream, as the last jet counteracts the main flow direction. This process reduces velocity variations, minimizing separation and eliminating discontinuities encountered by the flow as it converges toward the leading edge.

As a result, the flow acceleration leads to a substantial increase in the pressure differential between the front and rear sections of the airfoil, generating a significant lift enhancement. Simultaneously, the controlled deceleration downstream prevents excessive drag penalties, optimizing overall aerodynamic efficiency.

The specific scenario where the third jet is the primary blowing jet, while the other two jets counteract its effect with equal intensity, provides insight into why this strategy avoids discontinuities at high mass flow rates. Unlike the fixed-intensity cases analyzed in Section \ref{sec:intensityfixed}, where abrupt variations in flow properties were observed, the present DRL-based strategy achieves a smooth transition and maintains stability even under aggressive actuation conditions.

Figure \ref{fig:cp_B2} illustrates the pressure coefficient distribution along the airfoil, highlighting the impact of jet actuation on the flow field. The results indicate that the jets accelerate the flow between positions 0.2 and 0.3, followed by a reduction in velocity and a corresponding increase in pressure. However, this pressure increase is short-lived, and the flow quickly stabilizes without discontinuities. At the leading edge, both the upper and lower surfaces converge with nearly identical pressures, ensuring a smooth wake formation. This behavior is facilitated by the presence of two distinct layers on the upper surface: a thin low-velocity layer adjacent to the surface and a higher-velocity layer above it. As a result, the trailing edge encounter remains smooth, with the thicker layer slightly separating from the surface. However, the increased velocity at the airfoil's trailing edge relative to the inflow parabolic profile leads to an associated increase in pressure drag.

\begin{figure}[htbp!]
    \centering
    \begin{subfigure}[b]{0.8\linewidth}
        \centering
        \includegraphics[trim={0cm 0cm 1cm 0cm},clip,width=\linewidth]{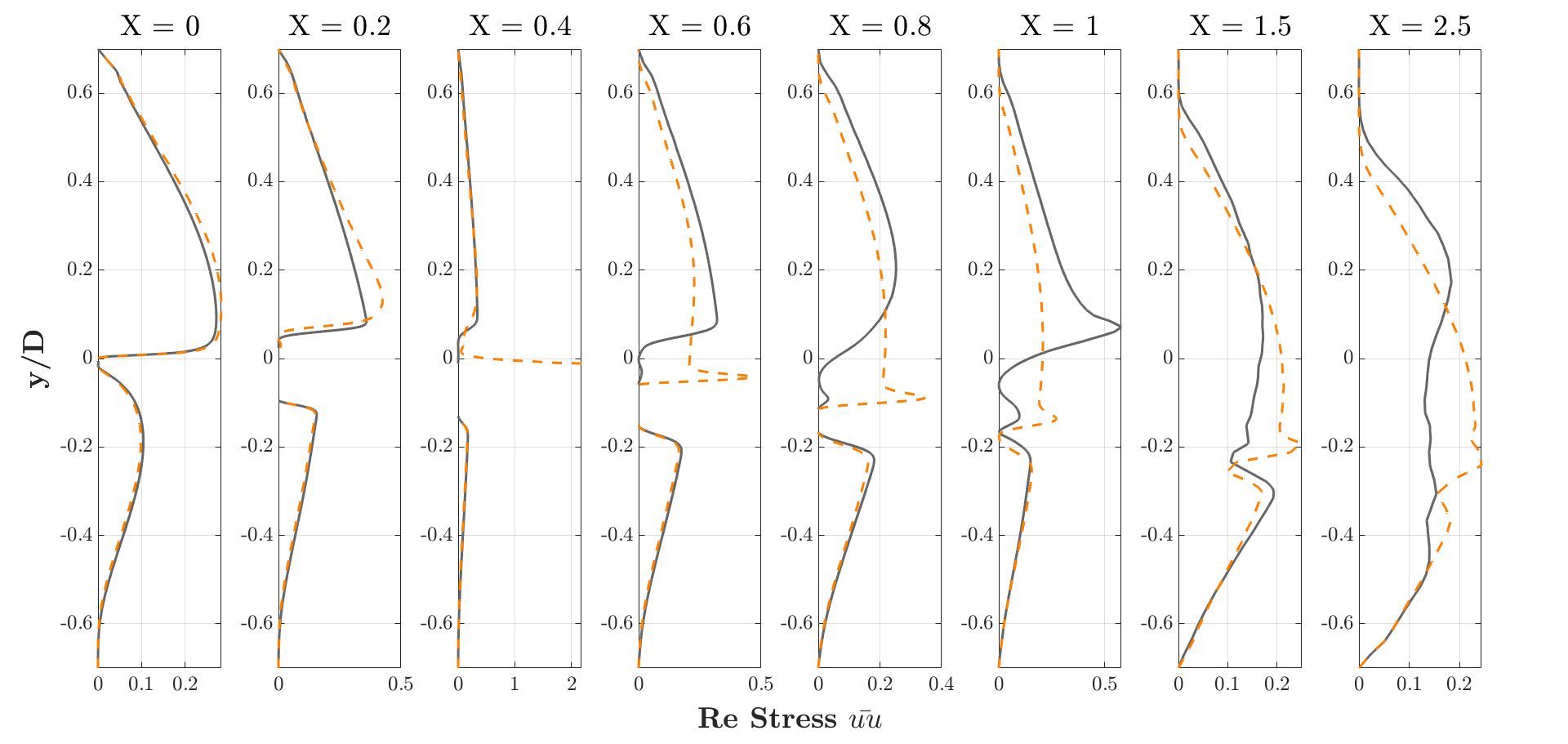}
        \caption{Reynolds stress \( \bar{uu} \).}
    \end{subfigure}
    \vfill
    \begin{subfigure}[b]{0.8\linewidth}
        \centering
        \includegraphics[trim={0cm 0cm 1cm 0cm},clip,width=\linewidth]{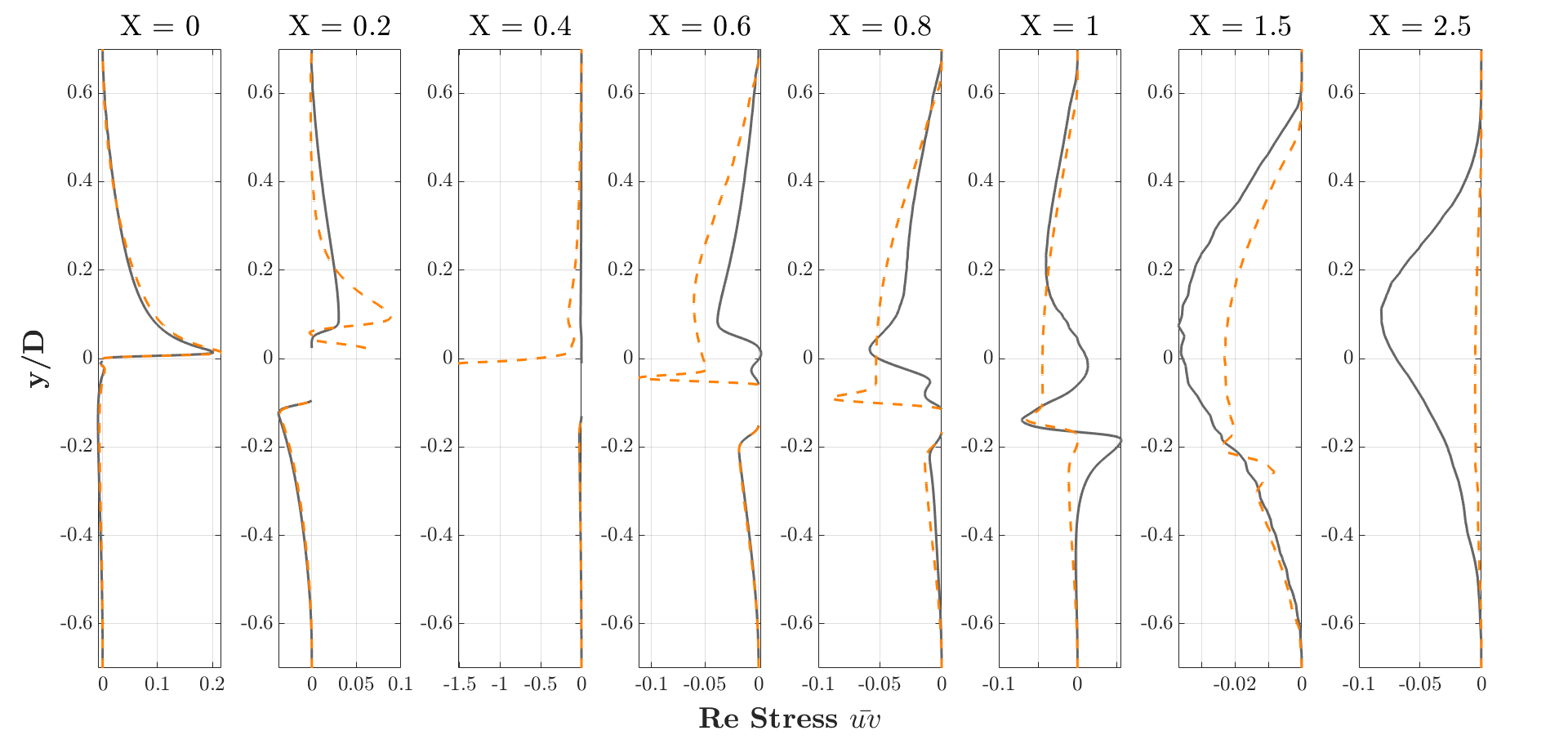}
        \caption{Reynolds stress \( \bar{uv} \).}
    \end{subfigure}
    \vfill
    \begin{subfigure}[b]{0.8\linewidth}
        \centering
        \includegraphics[trim={0cm 0cm 1cm 0cm},clip,width=\linewidth]{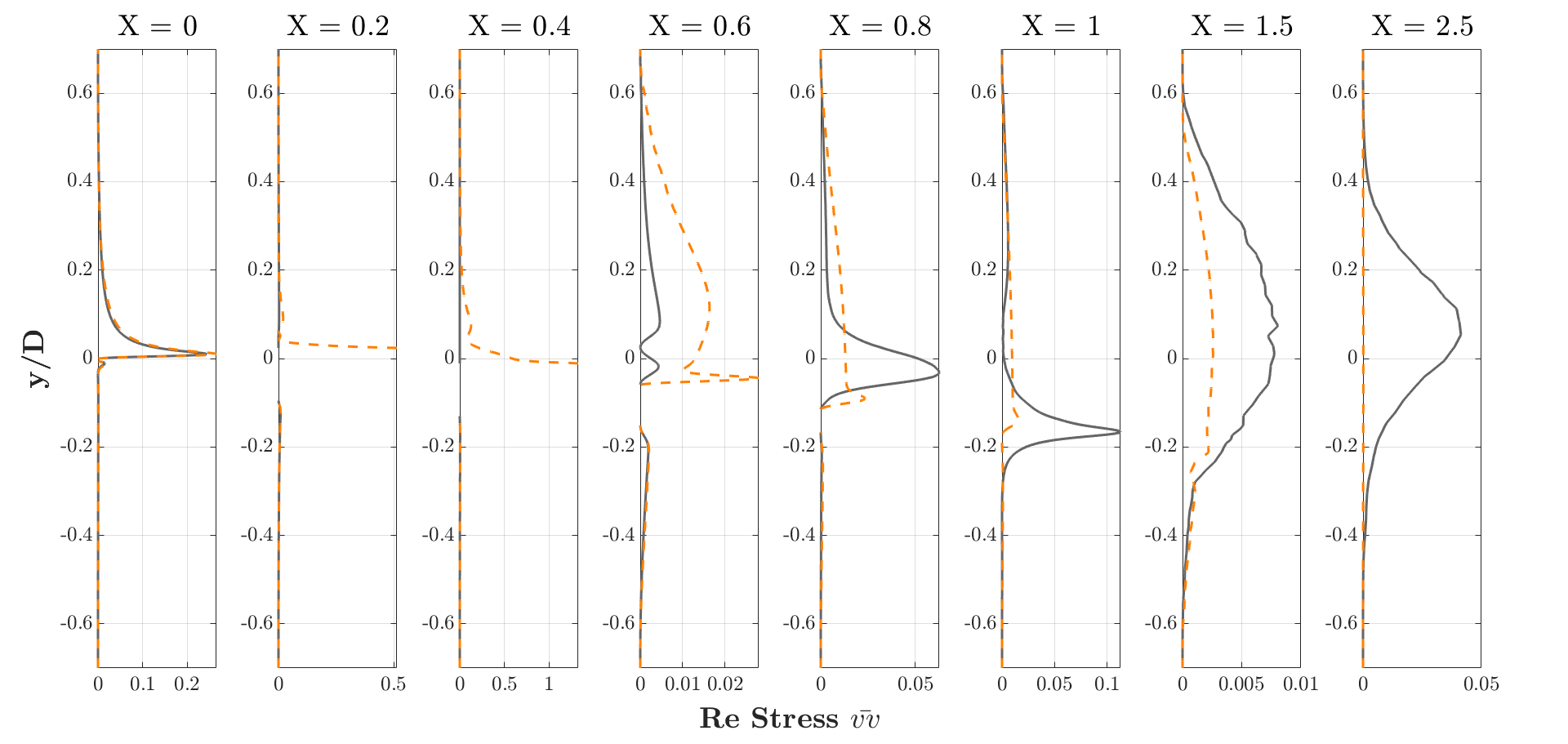}
        \caption{Reynolds shear stress \( \bar{vv} \).}
    \end{subfigure}
    \caption{Comparison of Reynolds stresses (\( \bar{uu} \), \( \bar{uv} \), and \( \bar{vv} \)) along the \( Y \)-coordinate at different streamwise locations (\( x/D \)). The dashed orange lines represent the actuated case, while the solid grey lines correspond to the baseline flow.}
    \label{fig:reynolds_stresses_combined_b2}
\end{figure}

The frequency spectrum confirms the absence of oscillations or dynamic behavior in the wake, reinforcing the steady nature of the flow. This can also be observed through the Reynolds stresses, as depicted in Figure \ref{fig:reynolds_stresses_combined_b2}.

Figure \ref{fig:reynolds_stresses_combined_b2}a reveals how the streamwise Reynolds stresses capture the two layers on the upper surface. The peak of the thicker layer begins after the jet region and extends into the wake. These peaks do not indicate a high-turbulence or separation region but rather an amplification of turbulence energy caused by external disturbances, in this case, the jets.

Figure \ref{fig:reynolds_stresses_combined_b2}b further illustrates the influence of jet actuation on the flow, as indicated by increased momentum transfer in the wall-normal direction. The previously mentioned flow layers are also visible here. When comparing the baseline separation region around \( X=0.8 \) with the steady actuated case, the wall-normal turbulence is significantly reduced and controlled. Additionally, at the trailing edge, no recirculation bubble is observed, as there are no signs of separation or discontinuities. This is further supported by the minimal stress levels at \( X=1 \), where both upper and lower surface flows merge smoothly. In the wake region, turbulence mixing in the wall-normal direction is absent at \( X=2.5 \), indicating the suppression of eddies.

\begin{figure}[htbp!]
\centering
\begin{subfigure}[b]{0.9\linewidth}
        \centering
        \includegraphics[trim={0 7cm 0 8cm}, clip, width=1\linewidth]{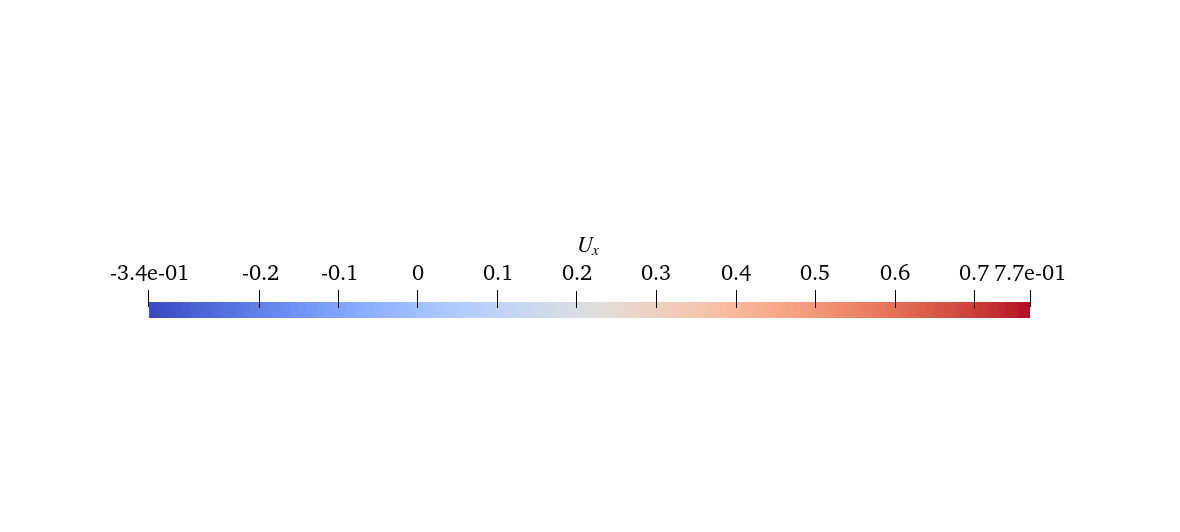}
    \end{subfigure}%
    \vfill
\begin{subfigure}[b]{1\linewidth}
    \centering
    \includegraphics[trim={0 5cm 0 5cm},clip,width=0.7\linewidth]{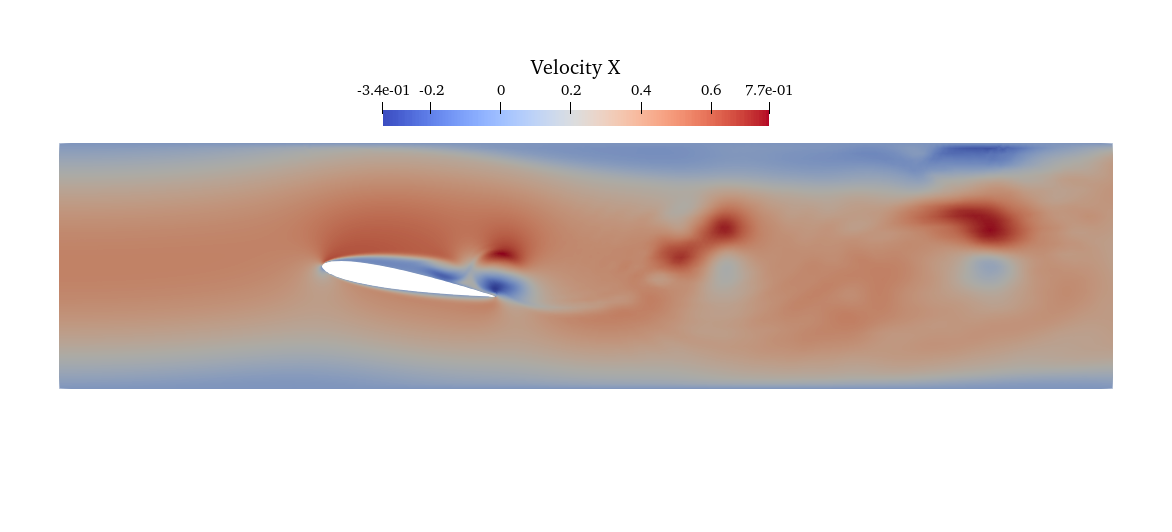}
    \caption{Baseline case.}
    \label{fig:low_intensity}
\end{subfigure}
\vfill
\begin{subfigure}[b]{1\linewidth}
    \centering
    \includegraphics[trim={0 5cm 0 5cm},clip,width=0.7\linewidth]{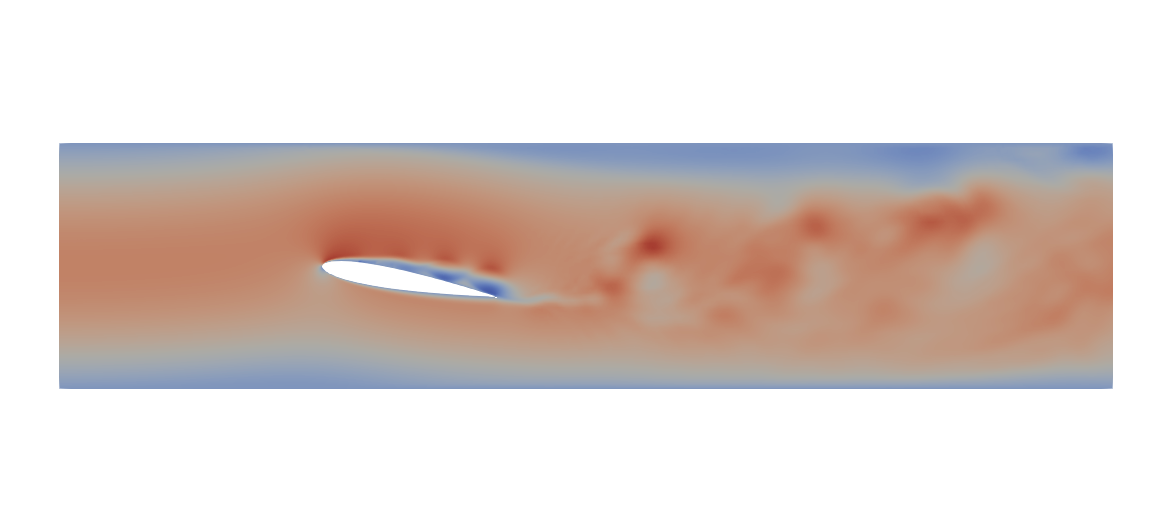}
    \caption{Drag reduction case.}
    \label{fig:medium_intensity}
\end{subfigure}
\vfill
\begin{subfigure}[b]{1\linewidth}
    \centering
    \includegraphics[trim={0 5cm 0 5cm},clip,width=0.7\linewidth]{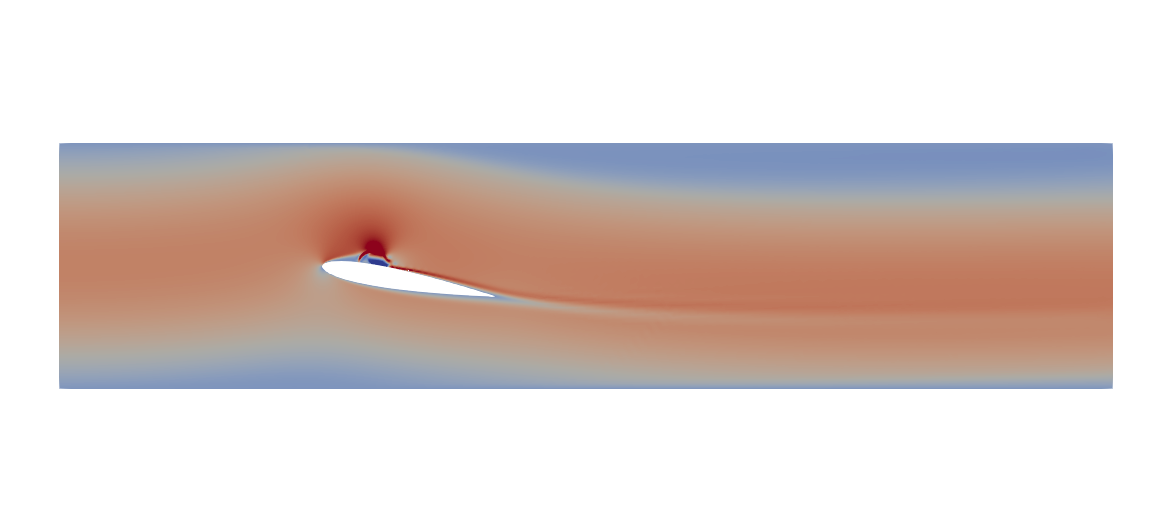}
    \caption{Efficiency enhancement case.}
    \label{fig:high_intensity}
\end{subfigure}
\caption{Comparison of the flow field in terms of streamwise velocity (\( U_x \)) for three different cases: (a) Baseline, showing natural flow behavior without actuation; (b) Drag reduction case, demonstrating the effects of DRL-based active flow control in minimizing drag, evidenced by reduced wake size and vortex shedding; and (c) Efficiency enhancement case, where actuation optimizes the flow for improved aerodynamic performance, resulting in a streamlined wake and enhanced boundary layer attachment. The color scale represents velocity magnitude in the \( x \)-direction.}
\label{fig:contours_final}
\end{figure}

Figure \ref{fig:reynolds_stresses_combined_b2}c depicts the Reynolds shear stresses, which highlight the formation of the recirculation bubble induced by the jets. The presence of this bubble is particularly noticeable at \( X=0.4 \), where substantial turbulence mixing and momentum transfer between the streamwise and wall-normal directions occur. The negative sign of the Reynolds shear stress suggests that this mixing is directed towards the wall, a phenomenon also observed in Figure \ref{fig:cp_B2}. Finally, at the wake profiles, no discernible eddies or discontinuities are present, indicating a significantly weaker wake compared to the baseline scenario. This can also be confirmed by examining Figure \ref{fig:contours_final}, where the wake is notably smoother compared to both the non-actuated and drag reduction cases.

\subsubsection{Comparison of the DRL results with periodic action strategies}

Various periodic-control jet flow strategies have been tested and compared against the optimized DRL-based strategies. Among these, the best-performing periodic case has been selected for direct comparison.

The results indicate that the optimal periodic strategy occurs when $\mathrm{jet}_1$ and $\mathrm{jet}_2$ operate in phase, while $\mathrm{jet}_3$ is phase-shifted by $180^\circ$. This strategy closely resembles the DRL-optimized strategy, as illustrated in Figure \ref{fig:actions_combined}, with the key distinction being that, in this case, the control actions follow a strictly periodic modulation at a frequency of $0.54$ Hz—three times the vortex shedding frequency. Among all tested periodic strategies, this approach has proven to be the most effective for drag reduction, achieving a 42\% decrease in drag. However, the DRL-optimized strategy outperforms it, yielding a 43.9\% reduction.

Regarding lift reduction, the periodic strategy results in a 32\% decrease, whereas the DRL strategy achieves a slightly greater reduction of 35\%.

\begin{figure}[htbp!]
    \centering
    \begin{subfigure}[b]{0.49\linewidth}
        \centering
        \includegraphics[trim={0cm 0cm 0cm 0cm},clip,width=\linewidth]{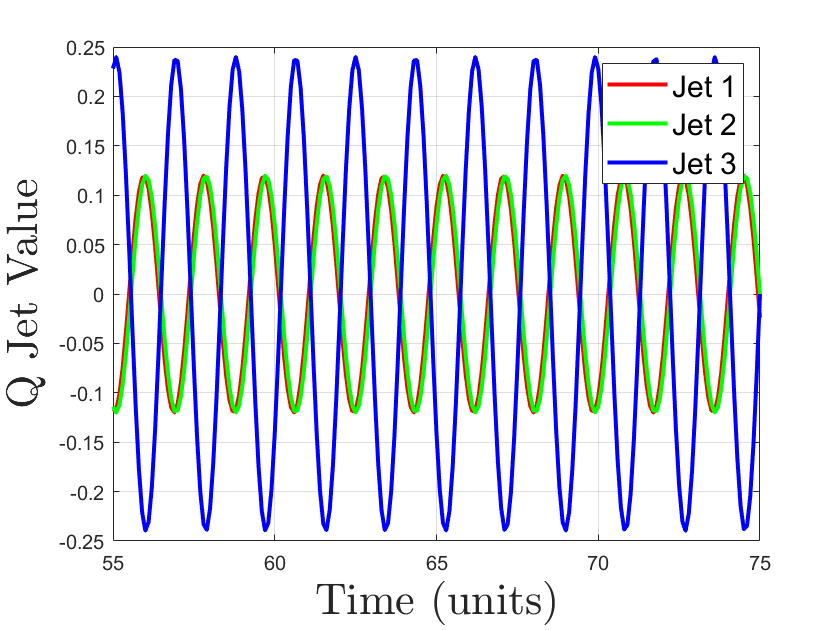}
        \caption{Jet actuation signals in the periodic control case.}
        \label{fig:actions_periodic}
    \end{subfigure}
    \hfill
    \begin{subfigure}[b]{0.49\linewidth}
        \centering
        \includegraphics[trim={0cm 0cm 0cm 0cm},clip,width=\linewidth]{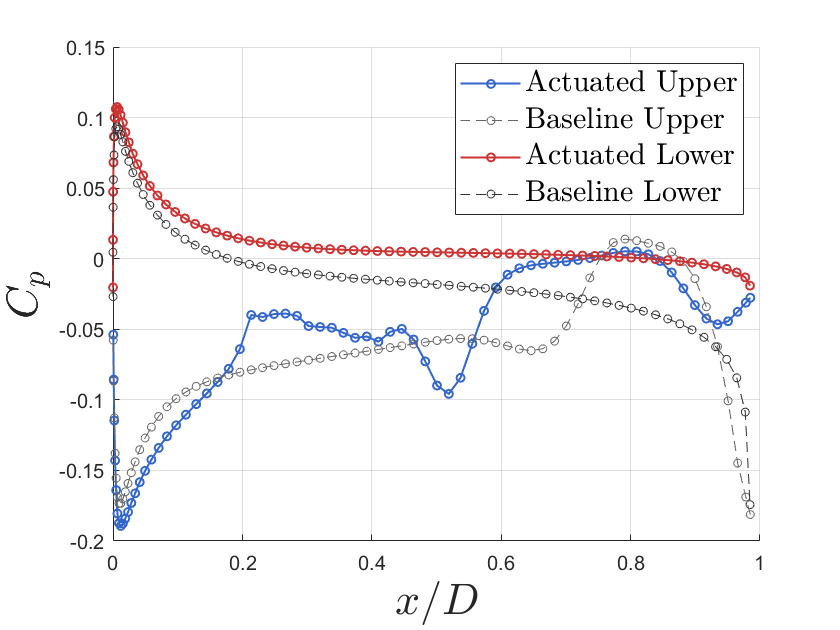}
        \caption{Pressure coefficient distribution along the airfoil for the periodic control case.}
        \label{fig:cp_periodic}
    \end{subfigure}
    \caption{Comparison of the pressure coefficient (\( C_p \)) along the airfoil for the baseline and periodic actuation cases.}
    \label{fig:actions_combined}
\end{figure}

It is important to note that the periodic strategy is less energy-efficient, requiring a 19.8\% higher mass flow rate to achieve slightly inferior aerodynamic performance. This underscores the superior capability of DRL in discovering optimized strategies that not only enhance aerodynamic performance but also improve energy efficiency. Furthermore, none of the periodic strategies tested achieved efficiency enhancements comparable to those identified in this study, emphasizing the significance of the DRL findings.

\section{\label{sec:conclusions}Conclusions}

This study has demonstrated the potential of deep reinforcement learning (DRL) for active flow control (AFC) in a two-dimensional NACA 0012 airfoil at a Reynolds number of 3000. The implementation of a DRL agent, capable of dynamic decision-making in real-time simulations, underscores its effectiveness in identifying optimal AFC strategies for both drag reduction and aerodynamic efficiency enhancement.

Initially, the DRL approach was designed to achieve significant aerodynamic drag reduction while incorporating additional regularization to maintain lift performance. This resulted in a 43.4\% decrease in drag; however, it also led to a 35.9\% reduction in lift, highlighting the inherent trade-off when prioritizing drag minimization. This outcome was primarily achieved through the suppression of vortex shedding and the mitigation of the recirculation bubble at the trailing edge. The DRL agent successfully identified dynamic strategies to weaken wake dynamics, resulting in a smoother downstream flow. The observed drag reduction highlights the adaptability and efficiency of reinforcement learning-based AFC. To achieve this reduction, the stabilization of the wake, characterized by decreased vortex shedding amplitudes, contributed to a reduction in pressure drag, while boundary layer modifications played a role in decreasing skin friction drag.

Conversely, when the focus shifted toward aerodynamic efficiency, the DRL agent exhibited its ability to optimally balance lift and drag, ultimately achieving a 58.6\% improvement in aerodynamic efficiency. Unlike the drag-reduction strategy, the efficiency-enhancement approach relied on higher jet intensities. The DRL agent leveraged this increased actuation to dynamically transition the flow from an unactuated vortex-shedding regime to a controlled, steady-state configuration. This efficiency improvement was driven by both an increase in lift and a reduction in drag, demonstrating that both performance metrics can be simultaneously optimized. The underlying physics involved the formation of a controlled closed recirculation bubble, which facilitated flow acceleration while simultaneously stabilizing the flow by smoothing the effective geometry encountered by the freestream.

When compared to fixed-intensity jet strategies, the DRL approach exhibited superior adaptability. While fixed policies were effective to some extent, they lacked the responsiveness necessary to optimize performance under dynamic conditions. In contrast, DRL-based strategies dynamically modulated actuation intensities based on real-time feedback, ensuring sustained aerodynamic improvements. Even in the efficiency-enhancement case, where jet actuation eventually stabilized at constant intensities, the DRL approach optimized the transition phase to minimize the time required to reach a steady-state flow. Furthermore, in the best-performing drag reduction cases, the DRL strategy achieved an additional 6\% drag reduction compared to fixed-intensity jets. Similarly, in efficiency-focused cases, DRL improved aerodynamic efficiency by 18\% by leveraging all three jets instead of just two, as employed in the fixed-intensity strategy. These findings highlight the superior performance of DRL in optimizing complex AFC systems.

In comparison to periodic-control strategies, the DRL approach achieved both superior performance and efficiency. While the most effective periodic actuation strategy reduced drag by 42\%, DRL further improved this to 43.9\%, requiring 19.8\% less mass flow rate. This highlights DRL’s ability to optimize aerodynamic performance while minimizing energy consumption, in contrast to periodic actuation strategies, which lack real-time adaptability to evolving flow conditions.

Despite these notable gains, the study encountered certain limitations and challenges. First, the channel configuration introduced ground effects, which led to augmented lift, potentially influencing direct comparisons with baseline studies. Second, a mass flow rate constraint was imposed to ensure numerical stability in the computational fluid dynamics (CFD) framework, which may have restricted the exploration of more aggressive actuation strategies.

Future research will focus on extending the DRL framework to higher Reynolds numbers and three-dimensional domains. Additionally, hybrid strategies combining DRL with traditional control methodologies will be investigated to enhance robustness. Further studies will also explore the scalability of DRL agents in optimizing flows with multiple objectives and complex constraints, aiming to refine AFC methodologies for practical aerodynamic applications.

\section*{Acknowledgments}

Ricardo Vinuesa acknowledges the financial support from ERCgrant no. 2021-CoG-101043998, DEEPCONTROL. Views and opinions expressed are however those of the author(s) only and do not necessarily reflect those of the European Union or the European Research Council. Neither the European Union nor the granting authority can be held responsible for them.

\nocite{*}

During the preparation of this work, the author(s) used Generative AI in order to enhance text clarity, improve formal language, and refine structure. After using this tool/service, the author(s) reviewed and edited the content as needed and take(s) full responsibility for the content of the publication.

\bibliography{aipsamp}

\end{document}